\documentclass[aps,prd,groupaddress,floatfix,tighten,nofootinbib,showpacs,amsfonts,superscriptaddress,a4paper,11pt]{revtex4}
\usepackage{url}
\usepackage{xcolor}
\usepackage{hyperref}
\usepackage{float}

\usepackage{changes}
\usepackage{feynmp}
\graphicspath{{./Diagrams/}}

\colorlet{shadecolor}{gray!15}
\definecolor{greenLinks}{rgb}{0,0.6,0}
\definecolor{blueLinks}{rgb}{0,0,0.6}
\definecolor{redLinks}{rgb}{0.6,0,0}
\definecolor{tempText}{rgb}{0.55,0.10,0.67}

\definecolor{eprintLinks}{rgb}{0.4,0.4,0.4}
\definecolor{journalLinks}{rgb}{0.6,0,0}

\DeclareMathSymbol{\not}{\mathrel}{symbols}{"36}
\usepackage{mathtools,amssymb,amsmath}
\usepackage{color}
\usepackage{soul}
\usepackage{ulem}
\usepackage{cancel}
\usepackage{graphicx,epsf,epsfig}
\usepackage{footnote}
\usepackage{lipsum}
\usepackage{multirow} 
\usepackage{tabulary}
\usepackage{rotating}
\usepackage{array}
\usepackage{color}
\usepackage{caption}
\usepackage{subcaption}

\def\slc#1{
 \setbox0=\hbox{$#1$}                      
 \dimen0=\wd0                              
 \setbox1=\hbox{/} \dimen1=\wd1            
 \ifdim\dimen0>\dimen1                     
 \rlap{\hbox to \dimen0{\hfil/\hfil}}      
 #1                                        
 \else                                     
 \rlap{\hbox to \dimen1{\hfil$#1$\hfil}}   
 /                                         
\fi}

\def\gs{\mathrel{\rlap{\raise 0.511ex \hbox{$>$}}{\lower 0.511ex \hbox{$\sim$}}}}
\def\ls{\mathrel{\rlap{\raise 0.511ex \hbox{$<$}}{\lower 0.511ex \hbox{$\sim$}}}}

\newcommand{\ba}{\begin{array}{c}}
\newcommand{\baz}{\begin{array}{cc}}
\newcommand{\barrr}{\begin{array}{rrr}}
\newcommand{\bad}{\begin{array}{ccc}}
\newcommand{\bav}{\begin{array}{cccc}}
\newcommand{\baf}{\begin{array}{ccccc}}
\newcommand{\bea}{\begin{equation} \begin{array}{c}}
\newcommand{\eea}{ \end{array} \end{equation}}
\newcommand{\ea}{\end{array}}

\newcommand{\br}{{\mathcal{BR}}}

\def\21{$\mathrm{SU(2)_L \otimes U(1)_Y}$ }

\newcommand{\ignore}[1]{}

\usepackage{float}
\usepackage{cleveref}
\newcommand{\bes}{\begin{subequations}}
\newcommand{\ees}{\end{subequations}}

\usepackage{slashed}

\newcommand{\olgafe}[1]{}
\newcommand{\soul}[1]{}

\definecolor{codegreen}{rgb}{0,0.6,0}
\definecolor{codegray}{rgb}{0.5,0.5,0.5}
\definecolor{codepurple}{rgb}{0.58,0,0.82}
\definecolor{backcolour}{rgb}{0.95,0.95,0.92}
\usepackage{listings}
\lstdefinestyle{mystyle}{
	backgroundcolor=\color{backcolour},   
	commentstyle=\color{blue},
	keywordstyle=\color{codegreen},
	numberstyle=\tiny\color{codegray},
	stringstyle=\color{red},
	basicstyle=\ttfamily\footnotesize,
	breakatwhitespace=false,         
	breaklines=true,                 
	captionpos=b,                    
	keepspaces=true,                 
	numbersep=5pt,                  
	showspaces=false,                
	showstringspaces=false,
	showtabs=false,                  
	tabsize=2
}

\begin{document}
\title{The General  One-loop Structure for the LFV Higgs Decays $H_r \to l_a l_b$ in multi-Higgs Models with Neutrino Masses}


\allowdisplaybreaks \allowdisplaybreaks[2]
 \newcommand{\AddrFCFMBUAP}{Fac. de Cs. F\'{\i}sico Matem\'aticas, 
  Benem\'erita Universidad Aut\'onoma de Puebla,\\
  Apdo. Postal 1152, Puebla, Pue.  72000, M\'exico.}
 \newcommand{\AddrFCEBUAP}{Fac. de Cs. de la Electr\'onica, 
  Benem\'erita Universidad Aut\'onoma de Puebla,\\
  Apdo. Postal 542, Puebla, Pue. 72000, M\'exico.}
\newcommand{\AddrCIFFU}{Centro Internacional de F\'{\i}sica Fundamental, 
	Benem\'erita Universidad Aut\'onoma de Puebla.}

\author{M. Zeleny-Mora}
 \email{moises.zeleny@alumno.buap.mx}
 \affiliation{\AddrFCFMBUAP}
  \affiliation{\AddrCIFFU}
  \author{J. Lorenzo D\'iaz-Cruz}
 \email{jldiaz@fcfm.buap.mx}
 \affiliation{\AddrFCFMBUAP}
 \affiliation{\AddrCIFFU}
\author{O. F\'elix-Beltr\'an}
 \email{olga.felix@correo.buap.mx}
 \affiliation{\AddrFCEBUAP}
 \affiliation{\AddrCIFFU}

\date{\today}

\begin{abstract}
	In this paper we present general formulae for the calculation of LFV Higgs decays 
	$H_r \to l_a l_b$ at one-loop, with $H_r$ being part of the Higgs spectrum of a generic multi-scalar extension of the 
	Standard Model (SM) with neutrino masses. We develop a method based on a classification of the particles appearing in the loop 
	diagrams (scalars, fermions and vectors), and by identifying  the corresponding couplings, we are able to present 
	compact expressions for the form factors involved in the amplitudes. Our results are applicable to models where Flavor Changing Neutral Currents (FCNC) are forbidden at tree-level, but change of flavor is induced by charged currents. 
	Then, as applications of our formalism,  we evaluate the branching ratio for the mode $h \to l_a l_b$, for two specific 
	models: the See-Saw Type I-$\nu$SM and the Scotogenic model (here $h = H_1$ corresponds to the SM-like Higgs boson);
	we find that the largest branching ratio for SM-like Higgs $h$ boson within the $\nu$SM  is of the order $\br(h\to \mu \tau)\simeq 10^{-12}$,
	while for the Scotogenic model we find $\br(h\to l_a l_b)\lesssim 10^{-9}$, which satisfy the latest experimental LHC results.
\end{abstract}

\maketitle

\newpage

\baselineskip 24pt

\clearpage

\setcounter{page}{1}

\maketitle

\section{Introduction\label{sec:intro}}
The minimal Standard Model (SM), which includes the linear realization of the Englert-Brout-Higgs (EBH) mechanism, has been confirmed at the Large Hadron Collider (LHC), thanks to the detection of a light Higgs  boson with $m_h=125.25 \pm 0.17$ GeV~\cite{Zyla:2020zbs}. Although its properties seem  consistent with the SM predictions~\cite{Chatrchyan:2012ufa,Aad:2012tfa}, one of the  goals of LHC is to test the Higgs properties as a search for possible signal of physics beyond the SM. 
So far, LHC has imposed strong limits on the scale of New Physics (NP)~\cite{Ellis2012,doi:10.1098/rsta.2011.0452,Wan2020}, but results in areas such as neutrinos and  Dark Matter (DM) suggest that some form of NP should exist~\cite{Bertone:2004pz}. In some of those SM extensions, one often has an extended Higgs sector~\cite{Cruz:2019vuo,2013,roig2016lepton,2017,2018,Arroyo-Urena:2020fkt}, which produces  distinctive signals that could  be tested at the LHC~\cite{ATLAS:2019pmk,Savina:2020mpk}.

NP models often include a Higgs boson with new features, for instance, besides SM deviations for
the Flavor-Conserving  (FC) Higgs-fermion couplings, it is possible to have Flavor-Violating (FV) Higgs-fermions interactions.
These FV Higgs couplings could arise at tree-level, as in the Two-Higgs Doublet Model (2HDM) of type III~\cite{Diaz:2002uk,DiazCruz:2004pj,Primulando:2019ydt,Ghosh:2021jeg}, or they could be induced  at loop-levels,  as in the Minimal SUSY SM (MSSM)~\cite{DiazCruz:2002er,Alvarado:2016par,PhysRevD.71.035011}. It is possible to explore those Lepton Flavor Violation (LFV) effects at low-energies, through the  LFV decays $l_b \to l_a\gamma$, $l_a \to l_b l_b l_c$~\cite{PhysRevD.71.035011,Kubo:2006yx,TheMEG:2016wtm,Nomura:2020dzw}. However, it is also possible to test these interactions by searching for the Higgs decays~\cite{DiazCruz:1999xe,DiazCruz:2004tr,Herrero:2009ad,Herrero:2017myd,Barradas-Guevara:2017ewn,Arana-Catania:2013xma,Hue:2015fbb,THAO2017159}. In particular,  the most recent search for LFV Higgs decays at LHC,  with center-of-mass energy $\sqrt{s} =13$ TeV, and an integrated luminosity  of $36.1$ fb$^{-1}$, has provided stronger bounds for the corresponding branching ratios. In particular, ATLAS reports $\br(h \to e \tau)=0.47\%(0.34^{0.13}_{-0.10})\%$, $\br(h \to \mu \tau)=0.28\%(0.37^{0.14}_{-0.10}\%)$~\cite{2020135069}, which are consistent with a zero value. In turn, CMS  reports (expected) upper limits on the production cross section times the branching fraction, which vary from 51.9 (57.4) fb to 1.6 (2.1) fb for the $\mu \tau$ and from 94.1 (91.6) fb to 2.3 (2.3) fb for the decay mode $e\tau$~\cite{Sirunyan_2020}.
  
The fact that the LHC has improved the methods to search for these LFV Higgs modes, and that the coming LHC phase will have higher luminosity, it will be possible to derive more restrictive bounds. This has motivated extra interest from the theoretical side, 
with multiple studies using this LFV Higgs signal to probe a variety of models, including: 
2HDM~\cite{Tsumura:2005mv,PhysRevD.73.016006,Primulando_2020, Vicente:2019ykr},  models with a low-scale flavon mixing with the SM-like Higgs boson~\cite{Arroyo-Urena:2018mvl}, models with 3-3-1 gauge symmetry~\cite{Hue:2015fbb}; See-Saw  model and its inverse version~\cite{PILAFTSIS199268,PhysRevD.47.1080,PhysRevD.71.035011,THAO2017159}, low scale See-Saw~\cite{PhysRevD.91.015001,PhysRevD.102.113006}, as well as  SUSY models, such as the MSSM~\cite{BRIGNOLE2003217,DiazCruz:2008ry} have been studied too. Other type of methods has been used to calculate the LFV Higgs decays, for example the Mass Insertion Approximation (MIA)~\cite{Arganda2016,PhysRevD.95.095029,10.3389/fphy.2019.00228} or Effective Field approach~\cite{PhysRevD.99.095040}

In this paper, we are interested in deriving general formulae for the calculation of the LFV Higgs decays $H_r \to \bar{l}_a l_b$ at one-loop level. Here $H_r$ (with $r =1,2,\dots$) denotes the Higgs bosons contained in multi-scalar extensions of the SM. We shall focus on models where change of flavor is induced by charged currents; FCNC associated with charged leptons and the neutral Higgs bosons $H_r$ are forbidden, but flavor violating neutrino-Higgs interactions are permitted.  We develop a method based on a classification of the particles appearing in the one-loop diagrams, which can be scalars, fermions, or vectors. Then, by identifying the corresponding couplings, one classifies all diagrams according to the number of fermions circulating in the loops.  Furthermore, as applications of our formalism, we evaluated  $\br(h \to l_a l_b)$  in the framework of two specific models: the See-Saw Type I-$\nu SM$, and the Scotogenic model where neutrino masses are  generated radiatively, and include a fermion DM candidate as well. Here,  $h=H_1$ is identified as the SM-like Higgs boson, with $m_h=125.25 \pm 0.17$ GeV. 

The organization of our paper goes as follows. Section~\ref{sec:higgsint} contains the generalities of our method, where the classification of couplings and Feynman diagrams are shown, as well as the resulting formulae for the  evaluation of LFV Higgs amplitude. Then, in Section~\ref{subsec:lfvhiggsseesaw} we start with a brief description of the model See-Saw Type I-$\nu SM$; we present the corresponding formulae for the calculations of $\br(h \to l_a l_b)$. Section~\ref{sec:lfvhiggsscot} contains the results for LFV Higgs decays in the Scotogenic model; it includes the formulae for the radiative neutrino masses, which are used as  constraints to obtain the allowed parameter space and to evaluate numerically the LFV Higgs branching ratios. Conclusions and perspectives are presented in Section~\ref{sec:concl}. Some formulae are left in Appendices~\ref{appen:feynmanrules} and~\ref{appen:loop-identities}, whereas Appendix~\ref{Appendix:Cancellation_divergences} contains some details about the cancellation of the divergences.

\section{Higgs's interactions and the one-loop structure of the decay $H_r \rightarrow l_a l_b$ \label{sec:higgsint}}

Flavor Changing Neutral Currents (FCNC) mediated by the Higgs boson are forbidden at tree-level in the SM, including the LFV Higgs decays. The 2HDM type III contains FCNC interactions  of the neutral scalars, which implies LFV Higgs  decays are allowed at tree-level. Here we are interested in models where FCNC mediated by neutral Higgs bosons are forbidden at tree-level (unless it involves non-SM particles), but can be induced by charged currents at one-loop level.

In order to discuss the one-loop LFV Higgs decay, we must first classify the Higgs interactions ($H_r$) with charged scalars  ($S^{\pm}$), fermions ($F^{0,\pm}$) and vectors ($V^{\pm}$). The Feynman rules are specified in the  Appendix~\ref{appen:feynmanrules}, and from there one we read the following factors associated with the corresponding interaction (written down in parentheses), namely:
\begin{eqnarray}\label{ec:VerticesGenericos1} \notag
c_r^{S^{\pm}} (H_r S^{\pm} S^{\mp}),  \\	 \notag
 c_r^{F^{\pm}} (H_r F^{\pm}F^{\mp}) , \, \,  c_{rL}^{F^0}  (H_r F_1^{0}F_2^{0}) ,  \, \,c_{rR}^{F^0} (H_r F_1^{0}F_2^{0}), \\
 c_r^{S^{+}V^{-}} (H_r S^{+} V^{-}) , \, \,  c_r^{S^{-}V^{+}} (H_r S^{-} V^{+}), 
	\, \,  c_r^{V^{\pm}} (H_r V^{\pm}_1 V^{\mp}_2).
\end{eqnarray}

In turn, we define the following factors for the interaction of charged scalars and vector bosons with the fermions circulating in the loop, namely:
\begin{align}\label{ec:VerticesGenericos2}\notag
	  c_L^{S^{\pm}} (F_1^0 F_2^{\mp} S^{\pm}),  \, \, &   \, \,c_R^{S^{\pm}} (F_1^0 F_2^{\mp} S^{\pm}) ,\\
	c_L^{V^{\pm}} (F_1^{0} F_2^{\mp} V^{\pm} ),  \, \, &   \, \, 
	c_R^{V^{\pm}} (F_1^{0} F_2^{\mp} V^{\pm} ).
\end{align}
As we can see from equation~\eqref{ec:VerticesGenericos1} assuming a generic interaction with fermions, {\it i.e.} $H_r F_1^0 F_2^0$, where $F_{1,2}^0$ denotes non-SM fermions, it is possible to have diagrams with two neutral fermions inside of loop, while from the other interactions with scalars and vectors we can only have one neutral fermion inside the loop.

All the couplings constants which represent interaction of $H_r$ are denoted with a index $r$, the subscripts $R$ and $L$ denote the quiral structure of the interaction, for example, the vertex associated to $F_1^0 F_2^{\pm} S^{\mp}$ has the structure $c_R^{S^{\pm}} P_R + c_L^{S^{\pm}} P_L$, with $P_{R,L}$ being the quirality projectors. 
\begin{figure}
  \begin{subfigure}[t]{0.3\linewidth}
  	\centering{
    \includegraphics[width=\linewidth]{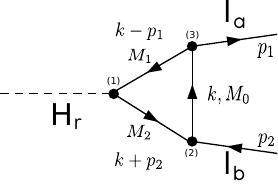}}
    \caption{}
    \label{fig:convencionestriangle}
  \end{subfigure}
  \begin{subfigure}[t]{.29\linewidth}
  		\centering{
    \includegraphics[width=\linewidth]{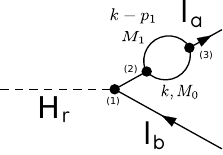}}
    \caption{}
    \label{fig:convencionesFX}
  \end{subfigure}
  \begin{subfigure}[t]{.3\linewidth}
  		\centering{
     \includegraphics[width=\linewidth]{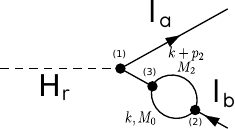}}
    \caption{}
    \label{fig:convencionesXF}
  \end{subfigure}
\caption{Conventions for momentum, labels of vertexes and particles appearing in the one-loop diagrams for $H_r \rightarrow l_a l_b$. In each diagram we put solid lines to represent all kind of particles in the loop and $k$ is the integration momentum. Vertexes are labeled with numbers inside of parenthesis as $(1)$,$(2)$ and $(3)$. Finally, the masses of particles $P_i$ in the loop are denoted by $M_i$, with $i = 0,1,2$.}
\label{fig:convencionesdiagramas}
\end{figure}

The general structure of the one-loop diagrams that contribute to the form factors $A^r_{L,R}$, Figure~\ref{fig:convencionesdiagramas}, showing the corresponding mass and momentum assignments, for each type of loop; Figure~\ref{fig:convencionestriangle} for triangles,~\ref{fig:convencionesFX} and~\ref{fig:convencionesXF} for bubbles. The amplitude for the LFV Higgs decay is written as:
\begin{equation}
i\mathcal{M} =  -i\overline{u}(p_1)(A^r_{L} P_L + A^r_{R} P_R )v(p_2). 
\end{equation}
Some relevant details of our calculation are summarized as follows:
\begin{itemize}	
	\item We used the 't Hooft-Feynman gauge, and work with dimensional regularization.
	
	\item We follow the Feynman rules for Dirac and Majorana fermions given by A. Denner et al.~\cite{DENNER1992467}.

	\item The one-loop amplitudes and the resulting form factors are expressed in terms of Passarino-Veltman (PV) functions, {\it i.e.} in terms of the scalar  integrals, as summarized in the Appendix~\ref{appen:loop-identities}. Each of the divergent one-loop integrals are divided into a divergent and convergent part, denoted as $\mathcal{F} = \text{Div}[\mathcal{F}] + f$, where $\mathcal{F}$ is a generic divergent PV function and $f$ its finite part.
		
	\item The analytical expressions for the resulting integrals are shown in the Appendix~\ref{appen:loop-identities}\footnote{These integrals have been obtained in Ref.~\cite{Hue:2015fbb}, and compared the reported results with those of Looptools~\cite{10.1093/ptep/ptw158}, finding good agreement.}.
		
	\item For numerical calculations, in contrast to Ref.~\cite{THAO2017159} that use approximate expressions for $f_0$ and $f_1$ functions, we consider the exact analytic expressions of these functions \eqref{ec:f01} derived from the definition of function $f_n$ \eqref{ec:fn} given in Appendix \ref{appen:loop-identities}. In this case, by means of mpmath library~\cite{mpmath} we can work with arbitrary precision and the stability of these functions is reached.
	
	\item  For handling of the amplitudes we created a simple implementation of the results given in the next Subsections~\ref{subsec:twoneufer}-~\ref{subsec:oneneufer} called \verb|OneLoopLFVHD|, which can be downloaded from the GitHub repository \href{https://github.com/moiseszeleny/OneLoopLFVHD}{OneLoopLFVHD}. 
\end{itemize}
Given the expression of the form factors, the decay width  $\Gamma(H_r \rightarrow l_a l_b)$  is given by
\begin{align}\label{ec:width}
\notag
\Gamma(H_r \rightarrow l_a l_b) &\equiv \Gamma(H_r \rightarrow l_a^{-} l_b^{+}) + \Gamma(H_r \rightarrow l_a^{+} l_b^{-}) \\ \notag
& = \frac{1}{8 \pi m_r} \left[1-\left( \frac{m_a^2 + m_b^2}{m_r} \right)^2  \right]^{1/2} \left[1-\left( \frac{m_a^2 - m_b^2}{m_r} \right)^2  \right]^{1/2} \\
&\times \left[ (m_r^2 -m_a^2 - m_b^2)(|A^{r}_L|^2 + |A^{r}_R|^2) - 4 m_a m_b \text{Re}(A^{r}_L A^{r\, *}_R) \right] ,
\end{align} 
where $m_r$, $m_a$ and $m_b$ are the masses associated with $H_r$, $l_a$ and $l_b$, respectively. We have taken into account the on-shell  conditions: $p_1^2 = m_a^2 $, $p_2^2 = m_b^2 $  and $p_r^2 \equiv (p_1 + p_2)^2 = m_r^2$.
\begin{figure}[t]
	\centering
	\begin{subfigure}[t]{.3\linewidth}
		\centering\includegraphics[width=.8\linewidth]{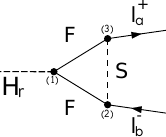}
		\caption{SFF}
		\label{dia:SFF}
	\end{subfigure}
	\begin{subfigure}[t]{.3\linewidth}
		\centering\includegraphics[width=.8\linewidth]{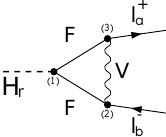}
		\caption{VFF}
		\label{dia:VFF}
	\end{subfigure}
	\caption{The two kinds of generic triangle diagrams with two fermions in the loop contributing to $H_r \to l_a l_b$.}
	\label{fig:TwoFermion}
\end{figure}
The specific diagrams with two fermions inside the loop are shown in Figure~\ref{fig:TwoFermion}, and are labeled according to the  internal loop particles. For example,  the triangle diagram (Figure~\ref{dia:SFF}) has two fermions and one scalar, and will be called SFF contribution. 
%

\subsection{Two neutral fermions inside of loop\label{subsec:twoneufer}}
Here and in the next sections, the  coupling constants $c^{(\kappa)}_{r,R,L}$
carry an upper index $\kappa$ ($\kappa=1,2,3$) to denote the site for each vertex in the diagram, following the conventions shown in Figure~\ref{fig:convencionesdiagramas}. For instance, in Figure~\ref{dia:SFF}, the vertex $H_r$FF is in site 1, the vertex FS$l_b$ in site 2, and the vertex FS$l_a$ is assigned to site 3.
We have just two diagrams with two fermions inside the loop (Figure~\ref{fig:TwoFermion}). The form factors for each one of these contributions are expressed as follows.

\subsubsection{SFF contribution\label{subsec:sffcont}}
Form factors associated with Figure~\ref{dia:SFF} have the following  generic structure:
\begin{subequations}\label{ec:FF-SFF}
	\begin{align}\notag
	A_R^r(SFF) &= c_{rL}^{F^0(1)} c_{R}^{S^\pm(2)} c_{R}^{S^\mp(3)} \mathcal{H}_1 + c_{rR}^{F^0(1)} c_{L}^{S^\pm(2)} c_{L}^{S^\mp(3)} \mathcal{H}_2 + c_{rL}^{F^0(1)} c_{L}^{S^\pm(2)} c_{R}^{S^\mp(3)} \mathcal{H}_3 \\ \notag
	& + c_{rR}^{F^0(1)} c_{R}^{S^\pm(2)} c_{L}^{S^\mp(3)} \mathcal{H}_4 +c_{rR}^{F^0(1)} c_{L}^{S^\pm(2)} c_{R}^{S^\mp(3)} \mathcal{H}_5 + c_{rL}^{F^0(1)} c_{R}^{S^\pm(2)} c_{L}^{S^\mp(3)} \mathcal{H}_6 \\
	& + c_{rR}^{F^0(1)} c_{R}^{S^\pm(2)} c_{R}^{S^\mp(3)} \mathcal{H}_7, \\ \notag
A_L^r(SFF) &=  c_{rR}^{F^0(1)} c_{L}^{S^\pm(2)} c_{L}^{S^\mp(3)} \mathcal{H}_1 + c_{rL}^{F^0(1)} c_{R}^{S^\pm(2)} c_{R}^{S^\mp(3)} \mathcal{H}_2 + c_{rR}^{F^0(1)} c_{R}^{S^\pm(2)} c_{L}^{S^\mp(3)} \mathcal{H}_3 \\ \notag
	& + c_{rL}^{F^0(1)} c_{L}^{S^\pm(2)} c_{R}^{S^\mp(3)} \mathcal{H}_4 +c_{rL}^{F^0(1)} c_{R}^{S^\pm(2)} c_{L}^{S^\mp(3)} \mathcal{H}_5 + c_{rR}^{F^0(1)} c_{L}^{S^\pm(2)} c_{R}^{S^\mp(3)} \mathcal{H}_6 \\ 
	& + c_{rL}^{F^0(1)} c_{L}^{S^\pm(2)} c_{L}^{S^\mp(3)} \mathcal{H}_7.
	\end{align}
\end{subequations}
The functions $\mathcal{H}_{k}$ ($k = 1,\dots,7$) can be expressed as
\begin{equation}
\begin{array}{lccl}
\mathcal{H}_{1}= X,&&&
\mathcal{H}_{2}= m_a m_b(\operatorname{C}_0 + \operatorname{C}_2 - \operatorname{C}_1),\\ 
\mathcal{H}_{3}= m_b M_2 \operatorname{C}_2,&& &
\mathcal{H}_{4}= m_a M_2(\operatorname{C}_0 - \operatorname{C}_1), \\ 
\mathcal{H}_{5}= m_b M_1 (\operatorname{C}_0 + \operatorname{C}_2), &&&
\mathcal{H}_{6}= -m_a M_1 \operatorname{C}_1 ,\\ 
\mathcal{H}_{7}=  M_1 M_2 \operatorname{C}_0, &&&
\end{array}
\label{ec:Hfunctions_SFF}
\end{equation}
where
\begin{equation}\label{ec:X}
    X =X(m_a,m_b,M_0,M_1,M_2) = \operatorname{B}_0^{(12)} + M_0^2 \operatorname{C}_0 + m_b^2 \operatorname{C}_2 - m_a^2 \operatorname{C}_1.
\end{equation}
For this type of diagram, we have that only $\mathcal{H}_1$ contains a divergent term, which is associated with $X$, and is given by $Div[\operatorname{B}_0^{(12)}]$.

In this work we follow the notation from~\cite{THAO2017159}, where $\operatorname{B}_{0,1}$ Passarino-Veltman (PV) functions are distinguished by the masses of the particles in the loop, namely, $B_{0,1}^{(1)} = B_{0,1}^{(1)}(M_0,M_1)$, $B_{0,1}^{(2)} = B_{0,1}^{(2)}(M_0,M_2)$ and $B_{0}^{(12)} = B_{0}^{(12)}(M_1,M_2)$. Similarly, $\operatorname{C}_{0,1,2} = \operatorname{C}_{0,1,2}(M_0, M_1, M_2)$.

To connect with the standard notations in the literature we have the following definitions:
\begin{equation}\label{ec:B_PV-Thao-Arganda}
	\begin{array}{lcl}
	B_{0}^{(1)} = B_{0}(m_{a}^2,M_0^2,M_1^2), & \qquad \qquad&
	B_{0}^{(2)} = B_{0}(m_{b}^2,M_0^2,M_2^2),\\ 
	B_{0}^{(12)} = B_{0}(M_1^2,M_2^2),& \qquad \qquad&
	B_{1}^{(1)} = -B_{1}(m_{a}^2,M_0^2,M_1^2),\\ 
	B_{1}^{(2)} = B_{1}(m_{b}^2,M_0^2,M_2^2),& \qquad \qquad&
	C_0 = C_0(0,0,m_r^2,M_0^2,M_1^2,M_2^2).
	\end{array}
\end{equation}
More details are found in the Appendix~\ref{appen:loop-identities}\footnote{The dependence on the masses for each PV function can be read from Figure~\ref{fig:convencionesdiagramas}, namely $M_0$ is between vertices 2 \& 3, $M_1$ is between vertices 3 \& 1, and $M_2$ is between 1 \& 2. The dependence of PV functions on the mass of $H_r$ is omitted in the notation $C_0 = C_0(m_r,M_0,M_1,M_2)$, but we use $C_0 = C_0(M_0,M_1,M_2)$. However, the dependence of the numerical expressions on $m_r$ is kept. For $C_0$ we take the approximation $m_a,m_b \rightarrow 0$.}
This specific notation allows us to omit the dependence of PV function for the coming expressions, considering the convention of Figure~\ref{fig:convencionesdiagramas} and identifying the $P_i$ particles inside the loop. For instance, in equation~\eqref{ec:Hfunctions_SFF} the dependence of $\operatorname{C}_{0,1,2} = \operatorname{C}_{0,1,2}(m_S,m_F,m_F)$ is in agreement with diagram~\ref{dia:SFF}, where $m_S$ is the scalar  mass and $m_F$ denotes the mass of the involves fermions.

\subsubsection{VFF contribution}
Next, the VFF contribution corresponds to the diagram of Figure~\ref{dia:VFF} and the corresponding form factors are given as follows
\begin{subequations}\label{ec:FF-VFF}
	\begin{align}\notag
	A_R^r(VFF) &= c_{rL}^{F^0(1)} c_{R}^{V^\pm(2)} c_{R}^{V^\mp(3)} \mathcal{G}_1 + c_{rR}^{F^0(1)} c_{L}^{V^\pm(2)} c_{L}^{V^\mp(3)} \mathcal{G}_2 + c_{rL}^{F^0(1)} c_{L}^{V^\pm(2)} c_{R}^{V^\mp(3)} \mathcal{G}_3 \\ \notag 
	&+ c_{rR}^{F^0(1)} c_{R}^{V^\pm(2)} c_{L}^{V^\mp(3)} \mathcal{G}_4 + c_{rR}^{F^0(1)} c_{R}^{V^\pm(2)} c_{R}^{V^\mp(3)} \mathcal{G}_5 + c_{rL}^{F^0(1)} c_{L}^{V^\pm(2)} c_{L}^{V^\mp(3)} \mathcal{G}_6 \\ 
	&+c_{rL}^{F^0(1)} c_{R}^{V^\pm(2)} c_{L}^{V^\mp(3)} \mathcal{G}_7, \\ \notag
	A_L^r(VFF) &= c_{rR}^{F^0(1)} c_{L}^{V^\pm(2)} c_{L}^{V^\mp(3)} \mathcal{G}_1 + c_{rL}^{F^0(1)} c_{R}^{V^\pm(2)} c_{R}^{V^\mp(3)} \mathcal{G}_2 + c_{rR}^{F^0(1)} c_{R}^{V^\pm(2)} c_{L}^{V^\mp(3)} \mathcal{G}_3\\ \notag 
	&+ c_{rL}^{F^0(1)} c_{L}^{V^\pm(2)} c_{R}^{V^\mp(3)} \mathcal{G}_4 + c_{rL}^{F^0(1)} c_{L}^{V^\pm(2)} c_{L}^{V^\mp(3)} \mathcal{G}_5 + c_{rR}^{F^0(1)} c_{R}^{V^\pm(2)} c_{R}^{V^\mp(3)} \mathcal{G}_6 \\ 
	&+c_{rR}^{F^0(1)} c_{L}^{V^\pm(2)} c_{R}^{V^\mp(3)} \mathcal{G}_7,
	\end{align}
\end{subequations}
 where the functions $\mathcal{G}_{k}$ (with $k = 1,..,7$), are
\begin{subequations}
	\begin{align}
	\mathcal{G}_{1} &= -(2-D)m_a M_2(\operatorname{C}_0 - \operatorname{C}_1),\\
	\mathcal{G}_{2} &= -(2-D)m_b M_2 \operatorname{C}_2,\\
	\mathcal{G}_{3} &= (D - 4) m_a m_b (\operatorname{C}_1 - \operatorname{C}_0 -\operatorname{C}_2),\\
	\mathcal{G}_{4} &= -(D X + 2(m_r^2 -m_a^2 - m_b^2)(\operatorname{C}_1 - \operatorname{C}_0 - \operatorname{C}_2)),\\
	\mathcal{G}_{5} &= (2-D)m_a M_1 \operatorname{C}_1,\\
	\mathcal{G}_{6} &= -(2-D)m_b M_1 (\operatorname{C}_0 + \operatorname{C}_2),\\
	\mathcal{G}_{7} &= -D M_1 M_2 \operatorname{C}_0.
	\end{align}
	\label{ec:Hfunctions_VFF}
\end{subequations}
We follow the usual convention for the integral dimension $D = 4-2\epsilon$ and $\epsilon\rightarrow 0$. In this case, the divergent term comes from $X$ function in $\mathcal{G}_{4}$ and is proportional to $Div[\operatorname{B}_0^{(12)}]$.

\subsection{One neutral fermion inside the loop\label{subsec:oneneufer}}
%
\begin{figure}[t]
	\centering
	\begin{subfigure}[t]{.3\linewidth}
		\centering\includegraphics[width=.8\linewidth]{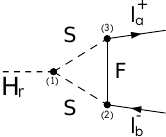}
		\caption{FSS}
		\label{dia:FSS}
	\end{subfigure}
	\begin{subfigure}[t]{.3\linewidth}
		\centering\includegraphics[width=.8\linewidth]{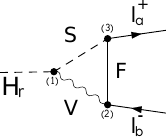}
		\caption{FSV}
		\label{dia:FSV}
	\end{subfigure}
	\begin{subfigure}[t]{.3\linewidth}
		\centering\includegraphics[width=.8\linewidth]{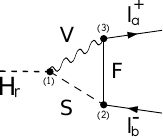}
		\caption{FVS}
		\label{dia:FVS}
	\end{subfigure}
	\medskip
	\begin{subfigure}[t]{.3\linewidth}
		\centering\includegraphics[width=.8\linewidth]{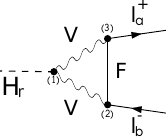}
		\caption{FVV}
		\label{dia:FVV}
	\end{subfigure}
	\begin{subfigure}[t]{.3\linewidth}
		\centering\includegraphics[width=.8\linewidth]{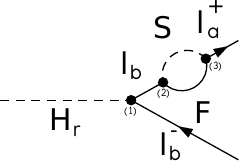}
		\caption{FS}
		\label{dia:FS}
	\end{subfigure}
	\begin{subfigure}[t]{.3\linewidth}
		\centering\includegraphics[width=.8\linewidth]{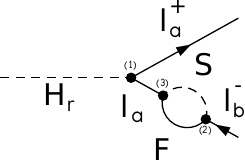}
		\caption{SF}
		\label{dia:SF}
	\end{subfigure}
	\medskip
	\begin{subfigure}[t]{.3\linewidth}
		\centering\includegraphics[width=.8\linewidth]{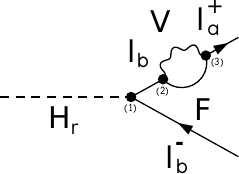}
		\caption{FV}
		\label{dia:FV}
	\end{subfigure}
	\begin{subfigure}[t]{.3\linewidth}
		\centering\includegraphics[width=.8\linewidth]{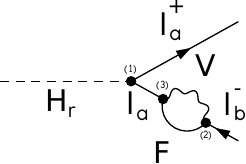}
		\caption{VF}
		\label{dia:VF}
	\end{subfigure}
	\caption{Generic diagrams with a single fermion in the one-loop contribution to $H_r \rightarrow l_a l_b$.}
	\label{fig:OneFermion}
\end{figure} 

Now, for diagrams with one fermion inside the loop, we have 8 contributions, denoted generically as $\Omega$: FSS, FSV, FVS, FVV, FS, SF, FV, VF; which are shown in Figure~\ref{fig:OneFermion} (see Table \ref{tab:OneFermionContributions})\footnote{
We follow the same conventions of diagrams SFF and VFF to label diagrams with one fermion inside the loop given in Figure~\ref{fig:OneFermion}.}.
Considering the generic couplings in equation~\eqref{ec:VerticesGenericos1} and the conventions on Figure~\ref{fig:convencionesdiagramas}, we found that the form factors for these contributions can be written as follows:
\begin{subequations}\label{ec:ALAR-OneFermion}
	\begin{align}\notag
	A^r_R(\Omega) &= m_{ab}^{-2} c^{I(1)}_r\left(c_R^{J(2)}c_R^{K(3)}\mathcal{H}_{RR}(\Omega) + c_L^{J(2)}c_L^{K(3)}\mathcal{H}_{LL}(\Omega) 
	\right.\\ & \qquad+ \left.  c_R^{J(2)}c_L^{K(3)}\mathcal{H}_{RL}(\Omega) + c_L^{J(2)}c_R^{K(3)}\mathcal{H}_{LR}(\Omega)\right),\\ \notag
	A^r_L(\Omega) &= m_{ab}^{-2} c^{I(1)}_r\left(c_L^{J(2)}c_L^{K(3)}\mathcal{H}_{RR}(\Omega) + c_R^{J(2)}c_R^{K(3)}\mathcal{H}_{LL}(\Omega) 
	\right.\\ & \qquad+ \left.  c_L^{J(2)}c_R^{K(3)}\mathcal{H}_{RL}(\Omega) + c_R^{J(2)}c_L^{K(3)}\mathcal{H}_{LR}(\Omega)\right).
	\end{align}
\end{subequations}
Here the  indices $I,J$ and $K$ denote the type of interaction in the $(i)$-th vertex, with $I = S^\pm, V^\pm, F^\pm, S^{\pm}V^\mp$, $J = S^\pm,V^\pm$ and $K = S^\mp,V^\mp$. Further, $m^2_{ab} = 1$ for triangle contributions and $m^2_{ab} = m_a^2 - m_b^2$ for bubble contributions; remember that $m_a$ and $m_b$ are the charged leptons masses. Finally, the functions $\mathcal{H}_{PQ}$ ($P,Q=R,L$)  are presented in Table~\ref{tab:OneFermionContributions}. 
\begin{center}
	{\small
		\begin{table}
			\centering
			\scalebox{0.95}{
				\begin{tabular}{||c|c|c|c|c||}
					\hline \hline
					\rule[-1ex]{0pt}{2.5ex} {$\Omega$\bf(Figure)} & $\mathcal{H}_{RR}$& $\mathcal{H}_{RL}$ & $\mathcal{H}_{LR}$&$\mathcal{H}_{LL}$ \\
					\hline \hline
					\rule[-1ex]{0pt}{2.5ex} FSS (\ref{dia:FSS})& $M_0 \operatorname{C}_0$ & $-m_b \operatorname{C}_2$ & $m_a \operatorname{C}_1$ & $0$ \\
					\hline
					\rule[-1ex]{0pt}{2.5ex} FSV (\ref{dia:FSV}) & $-X- 2m^2_{ar}\operatorname{C}_2 +m_a^2 \operatorname{C}_1$ & $-m_a M_0 (\operatorname{C}_1 -2 \operatorname{C}_0)$ & $-m_b M_0 (\operatorname{C}_0 - \operatorname{C}_2)$ & $-m_a m_b (\operatorname{C}_1 - 2 \operatorname{C}_2)$ \\
					\hline
					\rule[-1ex]{0pt}{2.5ex} FVS (\ref{dia:FVS}) & $m_a M_0 (\operatorname{C}_0 + \operatorname{C}_1)$ & $X - 2m^2_{br} \operatorname{C}_1 + m_b^2 \operatorname{C}_2$ & $-m_a m_b (\operatorname{C}_2 - 2 \operatorname{C}_1)$ & $-m_b M_0 (2 \operatorname{C}_0 + \operatorname{C}_2)$ \\
					\hline
					\rule[-1ex]{0pt}{2.5ex} FVV(\ref{dia:FVV}) & $-(D-2) m_a \operatorname{C}_1$ & $M_0 D \operatorname{C}_0$ & $0$ & $(D-2)m_b \operatorname{C}_2$ \\
					\hline
					\rule[-1ex]{0pt}{2.5ex} FS (\ref{dia:FS}) & $m_b M_0 \operatorname{B}_0^{(1)}$ & $m_a m_b \operatorname{B}_1^{(1)}$ & $m_a^2 \operatorname{B}_1^{(1)}$ & $m_a M_0 \operatorname{B}_0^{(1)}$ \\
					\hline
					\rule[-1ex]{0pt}{2.5ex} SF (\ref{dia:SF}) & $- m_a M_0 \operatorname{B}_0^{(2)}$ & $ m_b^2 \operatorname{B}_1^{(2)}$ & $m_a m_b \operatorname{B}_1^{(2)}$ & $- m_b M_0 \operatorname{B}_0^{(2)}$ \\
					\hline
					\rule[-1ex]{0pt}{2.5ex} FV (\ref{dia:FV}) & $(D-2) m_a m_b \operatorname{B}_1^{(1)}$ & $- D m_b M_0 \operatorname{B}_0^{(1)}$ & $- D m_a M_0 \operatorname{B}_0^{(1)}$ & $(D-2) m_a^2 \operatorname{B}_1^{(1)}$ \\
					\hline
					\rule[-1ex]{0pt}{2.5ex} VF (\ref{dia:VF}) & $(D-2) m_b^2 \operatorname{B}_1^{(2)}$ & $D m_a M_0 \operatorname{B}_0^{(2)}$ & $D m_b M_0 \operatorname{B}_0^{(2)}$ & $(D-2) m_a m_b \operatorname{B}_1^{(2)}$ \\
					\hline \hline
				\end{tabular}
			}
			\caption{Expression for the functions  $\mathcal{H}_{PQ}$ ($P,Q=R,L$) for each contribution $\Omega$ with a single fermion inside the loop. 
		}
			\label{tab:OneFermionContributions}
		\end{table}
	}
\end{center}
In our calculation, the neutrinos could be Dirac or Majorana, but it turns out that the difference in the corresponding Feynman rules does not have an effect in the final result. Namely, the Feynman rules for Dirac and Majorana change in the vertices, propagators and spinors. However, in the case of vertices, the charge conjugation transformations affect the  matrix $\gamma_\mu$, but the interaction vertex $V^{\pm} F_1^0 F_2^{\mp}$ preserves its structure (see Appendix~\ref{appen:feynmanrules}). For the propagators, we choose the direction in which both Dirac and Majorana propagators coincide. Finally, for the spinors, we consider that the final spinors are only Dirac leptons and the Feynman rules for Majorana fermions do not affect them (see Reference~\cite{DENNER1992467}).

\section{LFV Higgs Decays within the See-Saw Type I- $\nu$SM\label{subsec:lfvhiggsseesaw}}
As a first application of our formalism for LFV Higgs decays we have studied the See-Saw Type I- $\nu$SM. First computations to LFV Higgs decays in this context were performed in \cite{PILAFTSIS199268,PhysRevD.47.1080}, which were updated by Arganda et al.~\cite{PhysRevD.71.035011}; a most recent study was performed in ~\cite{THAO2017159}.
Lets us now discuss some details of the model, which will help us to evaluate the form factors.

\subsection{See-Saw Type I-$\nu $SM}
In the case of a See-Saw Type I-$\nu SM$ with three additional right-handed neutrinos, $N_{R, I} \sim(1,1,0)$, under $SU(3)_C \times SU(2)_L \times U(1)_Y$, the extended Lagrangian is 
\begin{equation}\label{ec:seesawLag}
-\Delta \mathcal{L}=Y_{\nu, a I} \overline{\psi_{L, a}} \widetilde{\phi} N_{R, I}+\frac{1}{2} \overline{\left(N_{R, I}\right)^{c}} M_{N, I J} N_{R, J}+\textrm{H. c.},
\end{equation}
where $a=1,2,3$; $I, J=1,2,3 $; $\psi_{L, a}=\left(\nu_{L, a}, l_{L, a}\right)^{T}$ are $S U(2)_{L}$ lepton doublets and $\left(N_{R, I}\right)^{c}=C \overline{N_{R, I}}^{T} .$ The Higgs doublet is given by $\phi=\left(G_{W}^{+},\left(h+i G_{Z}+v\right) / \sqrt{2}\right)^{T}$ with expectation value $\langle\phi\rangle=v/\sqrt{2}$, $v=246$ GeV and $\widetilde{\phi}=$ $i \sigma_{2} \phi^{*}$. 

Flavor states of active neutrinos are  denoted by $\nu_{L}=\left(\nu_{L, 1}, \nu_{L, 2}, \nu_{L, 3}\right)^{T}$, satisfying $\left(\nu_{L}\right)^{c} \equiv$ $\left(\left(\nu_{L, 1}\right)^{c},\left(\nu_{L, 2}\right)^{c},\left(\nu_{L, 3}\right)^{c}\right)^{T}$ and heavy right-handed neutrinos are  denoted by $N_{R}=\left(N_{R, 1}, N_{R, 2},  N_{R, 3}\right)^{T}$, with $\left(N_{R}\right)^{c}=\left(\left(N_{R, 1}\right)^{c},\left(N_{R, 2}\right)^{c}, \left(N_{R, 3}\right)^{c}\right)^{T}$. The neutrino mass term is given by 
\begin{equation}\label{ec:seesaw-mass}
-\mathcal{L}_{\mathrm{mass}}^{\nu} \equiv \frac{1}{2} \overline{\nu_{L}^{\prime}} \mathbf{M}^{\nu}\left(\nu_{L}^{\prime}\right)^{c}+\mathrm{H.c.}=\frac{1}{2} \overline{\nu_{L}^{\prime}}\left(\begin{array}{cc}
0 & \mathbf{M}_{D} \\
\mathbf{M}_{D}^{T} & \mathbf{M}_{N}
\end{array}\right)\left(\nu_{L}^{\prime}\right)^{c}+\mathrm{H.c.},
\end{equation}
where $\mathbf{M}_N$ is a symmetric and non-singular $3 \times 3$ matrix, $\mathbf{M}_D$ is a $3 \times 3$ matrix expressed as $(\mathbf{M}_D)_{a I} = Y_{\nu,aI}\langle \phi \rangle$. In the flavor basis, 
$\nu_{L}^{\prime} \equiv$ $\left(\nu_{L},\left(N_{R}\right)^{c}\right)^{T}$ and $\left(\nu_{L}^{\prime}\right)^{c}=\left(\left(\nu_{L}\right)^{c}, N_{R}\right)^{T}$. 
The matrix $\mathbf{M}^\nu$ is symmetric, therefore it can be diagonalized via $6 \times 6$ matrix $\mathbf{U}^\nu$, satisfying the unitary condition $\mathbf{U}^{\nu \dagger} \mathbf{U}^\nu = I$. Then,
\begin{equation}\label{ec:diagonalization}
\mathbf{U}^{\nu T} \mathbf{M}^\nu \mathbf{U}^\nu = \hat{\mathbf{M}}^\nu = \textrm{diag}(m_{n_1},m_{n_2},m_{n_3},m_{n_4},m_{n_5},m_{n_{6}}).
\end{equation}
$m_{n_i}$ ($i = 1,2,\ldots, 6$) are mass eigenvalues of the $6$ mass eigenstates $n_{L,i}$. In addition, the mixing matrix connects flavor and physical neutrinos by means of
\begin{equation}\label{ec:mixingU1}
\nu'_L = \mathbf{U}^{\nu *} n_L,\quad \text \quad (\nu'_L)^c = \mathbf{U}^{\nu} (n_L)^c,
\end{equation}
where $n_L \equiv (n_{L,1},n_{L,2},\ldots,n_{L,6})^T$. The physical Majorana neutrinos are $n_i\equiv (n_{L,i}, (n_{L,i})^c)^T = n_i^c = (n_i)^c$ ($i = 1,2,\ldots,6$). 
\begin{table}[H]
	\centering
	\scalebox{0.9}{
		\begin{tabular}{||c|c||c|c||}
			\hline 	\hline
			{\bf Vertex}&	{\bf Coupling}&	{\bf Vertex}&	{\bf Coupling}\\ \hline 
			$h W^{+\mu} W^{-\nu}$& $i g m_{W} g_{\mu \nu}$&$h G^{+} G^{-}$&$\frac{-igm_h^2}{2 m_W}$\\
			$h G^{+} {W}^{-\mu}$&$\frac{ig}{2}(p_{+}- p_0)_{\mu}$&$h G^{-} W^{+\mu}$&$\frac{i g}{2}\left(p_{0}-p_{-}\right)_{\mu}$\\
			$\bar{n}_{i} l_{a} W_{\mu}^{+}$&$\frac{i g}{\sqrt{2}} U_{a i}^{\nu} \gamma^{\mu} P_{L}$&$\overline{l_{a}} n_{j} W_{\mu}^{-}$&$\frac{i g}{\sqrt{2}} U_{a j}^{\nu *} \gamma^{\mu} P_{L}$\\
			$\bar{n}_{i} l_{a} G^{+}$&$-\frac{i g}{\sqrt{2} m_{W}} U_{a i}^{\nu}\left(m_{a} P_{R}-m_{n_i} P_{L}\right)$&$\overline{l_{a}} n_{j} G^{-}$&$-\frac{i g}{\sqrt{2} m_{W}} U_{a j}^{\nu *}\left(m_{a} P_{L}-m_{n_j} P_{R}\right)$\\
			$h\overline{n_i}n_j$&$\frac{-i g}{2 m_W}\left[C_{i j}\left(P_{L} m_{n_{i}}+P_{R} m_{n_{j}}\right) +C_{i j}^{*}\left(P_{L} m_{n_{j}}+P_{R} m_{n_{i}}\right)\right]$&$h\overline{l_a}l_a$&$\frac{-ig m_{a}}{2 m_W}$\\\hline \hline 
		\end{tabular}
	}\caption{Couplings involved on LFV Higgs decays for $\nu$SM model. Here, $C_{ij} = \sum_{c = 1}^3 U^\nu_{ci} U^{\nu *}_{cj}$. The $p_0$, $p_+$ and $p_-$ are incoming momenta of $h$, $G^+$, and $G^-$ respectively.}
\label{tab:couplings_seesaw}
\end{table}
Although we are interested on LFV Higgs decays, the radiative decays $l_a \to l_b \gamma$ are important allowed process by the inclusion of neutrino masses on the $\nu$SM, with strong experimental constraints (see Table~\ref{tb:bounds-lfv}). In the case of See-Saw models, the $\mathcal{BR}(l_a \to l_b \gamma)$ is given by~\cite{ILAKOVAC1995491}:
\begin{equation}\label{ec:BR_la_lbgamma}
\operatorname{BR}\left(l_{a} \rightarrow l_{b} \gamma\right)=\frac{\alpha_{W}^{3} s_{W}^{2}}{256 \pi^{2}}\left(\frac{m_a}{M_{W}}\right)^{4} \frac{m_a}{\Gamma_{l_{a}}}\left|G_{a b}\right|^{2},
\end{equation}
where $\Gamma_{l_{a}}$ is the total decay width of the lepton $l_a$, and
\begin{equation}
\begin{aligned}
G_{a b} &=\sum_{i=4}^{6} B_{a i}^{*} B_{b i} G_{\gamma}\left(\frac{m_{n_{i}}^{2}}{m_{W}^{2}}\right), \\
G_{\gamma}(x) &=-\frac{2 x^{3}+5 x^{2}-x}{4(1-x)^{3}}-\frac{3 x^{3}}{2(1-x)^{4}} \log x.
\end{aligned}
\end{equation}
Sum runs over the three heavy neutrinos $m_{n_{4,5,6}}$ and the mass of final lepton $m_{l_b} \to 0$.

%
\subsection{Form factors to $h \to l_a^+ l_b^-$}
The $\nu$SM with just one Higgs boson ($r \equiv h$), has contributions of ten diagrams to the form factors, as one can see in Table \ref{tab:seesaw_contributions}, where each diagram is summarized considering the particles $P_k$ with $k = 0,1,2$ involved in the loop.
Also, the form factors depend of the particles involved inside the loop as we shown in Section~\ref{sec:higgsint}, in this case $W^{\pm}$, $G^{\pm}$ and $n_i$. We define
\begin{equation}\label{ec:Dijab}
\Delta_{ij}^{ab} = \frac{g^{3}}{64 \pi^{2} m_{W}^{3}} U_{bj}^{\nu} U_{ai}^{\nu *}
\end{equation}
to obtain compact expressions for the form factors.
\begin{center}
	{
		\begin{table}
			\centering
			\begin{tabular}{||c|c|c|c|c||}
				\hline \hline
				\rule[-1ex]{0pt}{2.5ex}{\bf Number} & {\bf Diagram(\bf Figure)} & $P_0$ & $P_1$ & $P_2$ \\
				\hline \hline
				\rule[-1ex]{0pt}{2.5ex} 1& SFF(\ref{dia:SFF})& $G^{\pm}$ & $\overline{n}_i$ & $n_j$  \\
				\hline
				\rule[-1ex]{0pt}{2.5ex} 2& VFF(\ref{dia:VFF})& $W^{\pm}$ &$\overline{n}_i$ &  $n_j$  \\
				\hline
				\rule[-1ex]{0pt}{2.5ex} 3& FSS(\ref{dia:FSS})& $n_i$  &$G^{\pm}$ &  $G^{\mp}$  \\
				\hline
				\rule[-1ex]{0pt}{2.5ex} 4& FVS(\ref{dia:FVS}) & $n_i$  &$W^{\pm}$ &  $G^{\mp}$ \\
				\hline
				\rule[-1ex]{0pt}{2.5ex} 5& FSV(\ref{dia:FSV}) & $n_i$  &$G^{\pm}$ &  $W^{\mp}$  \\
				\hline
				\rule[-1ex]{0pt}{2.5ex} 6& FVV(\ref{dia:FVV}) & $n_i$  &$W^{\pm}$ &  $W^{\mp}$ \\
				\hline
				\rule[-1ex]{0pt}{2.5ex} 7& FV(\ref{dia:FV}) & $n_i$ &$W^{\pm}$ & --- \\
				\hline
				\rule[-1ex]{0pt}{2.5ex} 8 &FS(\ref{dia:FS}) & $n_i$  &$G^{\pm}$ &  ---  \\
				\hline
				\rule[-1ex]{0pt}{2.5ex} 9& VF(\ref{dia:VF}) & $n_i$  & ---  & $W^{\pm}$ \\
				\hline
				\rule[-1ex]{0pt}{2.5ex} 10& SF(\ref{dia:SF}) & $n_i$  & --- &  $G^{\pm}$  \\
				\hline \hline
			\end{tabular}
			\caption{Particles involved in each one-loop diagram that contribute to  $h \to l_a^+ l_b^-$ in the $\nu $SM model. Second column shows the structure (figure) of each diagram associated to the Figures~\ref{fig:TwoFermion} and~\ref{fig:OneFermion}. The remaining columns identify the particles $P_k$ inside the loop.}
			\label{tab:seesaw_contributions}
		\end{table}
	}
\end{center}
%

\subsubsection{Diagrams with two neutrinos inside the loop}
The interaction $h \overline{n}_i n_j$ allows diagrams with two neutrinos inside the loop; in first (second) diagram neutrinos interact with the charged leptons and $W^\pm$ ($G^{\pm}$) boson. In this regard, the  results of Section~\ref{subsec:twoneufer} are used to obtain the corresponding form factors. Then, the dependence of PV functions are given as $\operatorname{C_{0,1,2}} = \operatorname{C_{0,1,2}}{\left(m_W,m_{{n_i}},m_{{n_j}} \right)}$  and $\operatorname{B_{0}^{(12)}} =\operatorname{B_{0}^{(12)}} {\left(m_{{n_i}},m_{{n_j}} \right)}$. Then, interaction $h \overline{n}_i n_j$ appears in two diagrams: 1) SFF structure and 2) VFF structure (see Table~\ref{tab:seesaw_contributions}).

\bigskip

{\it Diagram 1:}  SFF structure

For the diagram with SFF structure,  the couplings constants can be extracted from Table~\ref{tab:couplings_seesaw} as follows
\begin{equation}
\begin{aligned}
c_{hR}^{F^0 (1)}(h\overline{n}_i n_j) &= -\frac{ig}{2 m_W}(m_{n_j} C_{ij} + m_{n_i} C_{ij}^*),
&c_{hL}^{F^0 (1)}(h\overline{n}_i n_j) &= -\frac{ig}{2 m_W}(m_{n_i} C_{ij} + m_{n_j} C_{ij}^*),\\
c_R^{S^\pm (2)}(\overline{n}_j G^+ l_b^-) &= -\frac{ig}{\sqrt{2} m_W}m_{b} U_{bj}^{\nu}, 
&c_L^{S^\pm (2)}(\overline{n}_j G^+ l_b^-) &= \frac{ig}{\sqrt{2} m_W}m_{n_i} U_{bj}^{\nu},\\
c_R^{S^\pm (3)}(n_i G^- l_a^+) &= \frac{ig}{\sqrt{2} m_W}m_{n_i} U_{ai}^{\nu *}, 
&c_L^{S^\pm (3)}(n_i G^- l_a^+)  &= -\frac{ig}{\sqrt{2} m_W}m_{b} U_{ai}^{\nu *}.\\
\end{aligned}
\end{equation}
Then, using equation~\eqref{ec:FF-SFF} the form factors are given by 
\begin{equation}
	\begin{array}{rcl}
A_L^h(G^{\pm}\overline{n}_i n_j)_{ab} &=&  {m}_{a}  \left[\left(\left(\operatorname{{{B^{(12)}_{0}}}} + m_{W}^{2} \operatorname{C_{0}}\right) m_{n_j}^{2} - \left({m}_{a}^{2} m_{n_j}^{2} + {m}_{b}^{2} m_{n_i}^{2} - 2 m_{n_i}^{2} m_{n_j}^{2}\right) \operatorname{C_{1}} \right) {C}_{ij}  \right.\\ &+& \left. 
 \left(\operatorname{{{B^{(12)}_{0}}}} + m_{W}^{2} \operatorname{C_{0}} - \left({m}_{a}^{2} + {m}_{b}^{2} - m_{n_i}^{2} - m_{n_j}^{2}\right) \operatorname{C_{1}} \right) C^*_{ij} m_{n_i} m_{n_j}\right] \Delta_{ij}^{ab} , \\ 
A_R^h(G^{\pm}\overline{n}_i n_j)_{ab} & = & {m}_{b} \left[\left(\left(\operatorname{{{B^{(12)}_{0}}}} + m_{W}^{2} \operatorname{C_{0}}\right) m_{n_i}^{2} + \left({m}_{a}^{2} m_{n_j}^{2} + {m}_{b}^{2} m_{n_i}^{2} - 2 m_{n_i}^{2} m_{n_j}^{2}\right) \operatorname{C_{2}} \right) {C}_{ij} \right.\\&+ & \left.  \left(\operatorname{{{B^{(12)}_{0}}}} + m_{W}^{2} \operatorname{C_{0}} + \left({m}_{a}^{2} + {m}_{b}^{2} - m_{n_i}^{2} - m_{n_j}^{2}\right) \operatorname{C_{2}} \right) C^*_{ij} m_{n_i} m_{n_j} \right]  \Delta_{ij}^{ab}.
\end{array}
\end{equation}

\bigskip

{\it Diagram 2:}  VFF structure

In this case we have a diagram with VFF structure, the couplings constants of vertex $(1)$ are the same as diagram 1. However, the coupling constants for vertices (2) and (3) are given by
\begin{equation}
\begin{aligned}
c_R^{V^\pm (2)}(\overline{n}_j W^+ l_b^-) &= 0; 
&c_L^{V^\pm (2)}(\overline{n}_j W^+ l_b^-) &= \frac{ig}{\sqrt{2}} U_{bj}^{\nu};\\
c_R^{V^\pm (3)}(n_i W^- l_a^+) &= 0; 
&c_L^{V^\pm (3)}(n_i W^- l_a^+) &= \frac{ig}{\sqrt{2}} U_{ai}^{\nu *}.\\
\end{aligned}
\end{equation}
Following the results for VFF contribution shown in equation~\eqref{ec:FF-VFF}, we have
\begin{equation}
	\begin{array}{rcl}
A_L^h(W^{\pm}\overline{n}_i n_j)_{ab} &=& 2 m_{W}^{2} {m}_{a}  \left(\left(\left(m_{n_i}^{2} + m_{n_j}^{2}\right) \operatorname{C_{1}} - \operatorname{C_{0}} m_{n_j}^{2}\right) {C}_{ij} - \left(\operatorname{C_{0}} - 2 \operatorname{C_{1}}\right) C^*_{ij} m_{n_i} m_{n_j}\right)  \Delta_{ij}^{ab}, \\
A_R^h(W^{\pm}\overline{n}_i n_j)_{ab} & =&- 2 m_{W}^{2} {m}_{b}  \left(\left(\left(m_{n_i}^{2} + m_{n_j}^{2}\right) \operatorname{C_{2}} + \operatorname{C_{0}} m_{n_i}^{2}\right) {C}_{ij} + \left(\operatorname{C_{0}} + 2 \operatorname{C_{2}}\right) C^*_{ij} m_{n_i} m_{n_j}\right) \Delta_{ij}^{ab} . 
\end{array}
\end{equation}
%

\subsubsection{Diagram with one neutrino inside the loop}

The remaining diagrams are obtained following the result of  Section~\ref{subsec:oneneufer}. In this matter, the dependence of PV functions is $\operatorname{C_{0,1,2}} = \operatorname{C_{0,1,2}}{\left(m_{{n_i}},m_{{W}},m_{{W}} \right)}$, $\operatorname{B_{t}^{(s)}} =\operatorname{B_{t}^{(s)}} {\left(m_{{n_i}},m_{{W}} \right)}$ and $\operatorname{B_{0}^{(12)}}  =\operatorname{B_{0}^{(12)}} {\left(m_{{W}},m_{{W}} \right)}$ with  $t = 0,1$  and $s = 1,2$. Following the conventions of Figure~\ref{fig:convencionesdiagramas}, for the one fermion contributions we have the following possible structures XFF, FX, XF, FXY, where X, Y can be $G^{\pm}$ or $W^{\pm}$ (X $\neq$ Y), then, the coupling constants $c_{R,L}^{J(i)}$ for $i =2,3$, can be taken for the coupling constant of diagrams 1 and 2 depending on X and Y. However, the coupling constant $c^{I(1)}_h$ will be different for each diagram.

\bigskip

{\it Diagram 3:}  FSS structure

We have a diagram with FSS structure, and the coupling constant is $c_h^{S^\pm (1)}(h G^- G^+) = -\frac{ig m_h^2}{2 m_W}$. Then considering Table~\ref{tab:OneFermionContributions}, we have
\begin{equation}
\begin{array}{rcl}
A_L^h(n_i G^- G^+)_{ab} & = &{m}_{h}^{2} m_{a} \left(\left( \operatorname{C_{0}}  - \operatorname{C_{1}} \right) m_{n_i}^{2} + \operatorname{C_{2}} {m}_{b}^{2}\right) \Delta_{ii}^{ab},  \\
A_R^h(n_i G^- G^+)_{ab} & = &{m}_{h}^{2} m_{b} \left(\left( \operatorname{C_{0}} +  \operatorname{C_{2}} \right) m_{n_i}^{2} - \operatorname{C_{1}} {m}_{a}^{2}\right) \Delta_{ii}^{ab}. 
\end{array}
\end{equation}

\bigskip

{\it Diagrams 4 and 5:} FVS and FSV structures

The diagrams 4 and 5 have the FVS and FSV structures respectively, and the coupling constants for the first vertex are given by $c_h^{S^+ V^- (1)}(h W^- G^+) = c_h^{S^- V^+ (1)}(h G^- W^+) = \frac{ig}{2}$,
\begin{equation}
	\begin{array}{rcl}
A_L^h(n_i W^- G^+)_{ab} &=& m_{W}^{2}{m}_{a} \left[\left(2 {m}_{b}^{2} - m_{n_i}^{2}\right) \operatorname{C_{1}} - \operatorname{C_{0}} m_{n_i}^{2} - \operatorname{C_{2}} {m}_{b}^{2}\right] \Delta_{ii}^{ab}, \\ 
A_R^h(n_i W^- G^+)_{ab} &=&  m_{W}^{2} {m}_{b} \left[\operatorname{{{B^{(12)}_{0}}}} + \left(2 {m}_{b}^{2} + m_{n_i}^{2}\right) \operatorname{C_{2}} - \left({m}_{a}^{2} + 2 {m}_{b}^{2} - 2 {m}_{h}^{2}\right) \operatorname{C_{1}} + 3 \operatorname{C_{0}} m_{n_i}^{2} \right] \Delta_{ii}^{ab}, \\
A_L^h(n_i G^- W^+)_{ab} & =&  m_{W}^{2} m_{a}m_{n_i}  \left(\operatorname{{{B^{(12)}_{0}}}}- \left(2 {m}_{a}^{2} + m_{n_i}^{2}\right) \operatorname{C_{1}} + \left(2 {m}_{a}^{2} + {m}_{b}^{2} - 2 {m}_{h}^{2}\right) \operatorname{C_{2}} + 3 \operatorname{C_{0}} m_{n_i}^{2} \right)  \Delta_{ii}^{ab},   \\
A_R^h(n_i G^- W^+)_{ab} & =&  m_{W}^{2} {m}_{b} m_{n_i}  \left(- \left(2 {m}_{a}^{2} - m_{n_i}^{2}\right) \operatorname{C_{2}} - \operatorname{C_{0}} m_{n_i}^{2} + \operatorname{C_{1}} {m}_{a}^{2}\right) \Delta_{ii}^{ab} .
	\end{array}
\end{equation}

\bigskip

{\it Diagram 6:} FVV structure

In this case we have a diagram with FVV structure, arising from the electroweak interaction with the $W$ boson, $c_h^{V^\pm (1)}(hW^- W^+) = i g m_W$. Then, following the results for the FVV contribution shown in  Table~\ref{tab:OneFermionContributions}, we obtain:
\begin{equation}
\begin{array}{rcl}
A_L^h(n_i W^- W^+)_{ab} & = - 4 m_{W}^{4} {m}_{a} \operatorname{C_{1}}\Delta_{ii}^{ab},\\
A_R^h(n_i W^- W^+)_{ab} & =  4 m_{W}^{4}  {m}_{b} \operatorname{C_{2}}\Delta_{ii}^{ab}.
\end{array}
\end{equation}

\bigskip

{\it Diagrams 7-10:} FX and XF (X= $G^{\pm}$ or $W^{\pm}$ ) structures

We have diagrams with FX and XF structures, and the couplings constant $c_h^{F^\pm (1)}(h l_k^- l_k^+) = -\frac{ig m_k}{2 m_W}$ with $k=b$ ($k=a$) for structures FX (XF).

Using the results of Table~\ref{tab:OneFermionContributions} for FV contribution, we have
\begin{equation}
	\begin{array}{rcl}
A_L^h(n_i W^{\pm})_{ab} & = &- \frac{2 m_{W}^{2}  {m}_{a} {m}_{b}^{2} }{{m}_{a}^{2} - {m}_{b}^{2}} \operatorname{{{B^{(1)}_{1}}}}\Delta_{ii}^{ab} , \\
A_R^h(n_i W^{\pm})_{ab} & = &- \frac{2 m_{W}^{2}  {m}_{a}^2 {m}_{b} }{{m}_{a}^{2} - {m}_{b}^{2}} \operatorname{{{B^{(1)}_{1}}}}\Delta_{ii}^{ab}, \\
A_L^h(n_i G^{\pm})_{ab} & =  & \frac{{m}_{a} {m}_{b}^2 }{{m}_{a}^{2} - {m}_{b}^{2}}  \left(- \left({m}_{a}^{2} + m_{n_i}^{2}\right) \operatorname{{{B^{(1)}_{1}}}} + 2 \operatorname{{{B^{(1)}_{0}}}} m_{n_i}^{2}\right) \Delta_{ii}^{ab},\\
A_R^h(n_i G^{\pm})_{ab} & =  &\frac{ {m}_{b} }{{m}_{a}^{2} - {m}_{b}^{2}} \left(\left({m}_{a}^{2} + {m}_{b}^{2}\right) \operatorname{{{B^{(1)}_{0}}}} m_{n_i}^{2} - \left({m}_{b}^{2} + m_{n_i}^{2}\right) \operatorname{{{B^{(1)}_{1}}}} {m}_{a}^{2}\right)\Delta_{ii}^{ab}, \\
A_L^h(W^{\pm} n_i)_{ab} & = &- \frac{2 m_{W}^{2}  {m}_{a} {m}_{b}^{2} }{ {m}_{a}^{2} - {m}_{b}^{2}} \operatorname{{{B^{(2)}_{1}}}}\Delta_{ii}^{ab},\\
A_R^h(W^{\pm} n_i)_{ab} & = &- \frac{2 m_{W}^{2}  {m}_{a}^{2} {m}_{b} }{ {m}_{a}^{2} - {m}_{b}^{2}} \operatorname{{{B^{(2)}_{1}}}}\Delta_{ii}^{ab},\\
A_L^h(G^{\pm} n_i)_{ab} & = &- \frac{ {m}_{a}}{{m}_{a}^{2} - {m}_{b}^{2}}  \left(\left({m}_{a}^{2} + {m}_{b}^{2}\right) \operatorname{{{B^{(2)}_{0}}}} m_{n_i}^{2} + \left({m}_{a}^{2} + m_{n_i}^{2}\right) \operatorname{{{B^{(2)}_{1}}}} {m}_{b}^{2}\right) \Delta_{ii}^{ab}, \\
A_R^h(G^{\pm} n_i)_{ab} & = &- \frac{ {m}_{a}^2 {m}_{b}}{{m}_{a}^{2} - {m}_{b}^{2}} \left(\left({m}_{b}^{2} + m_{n_i}^{2}\right) \operatorname{{{B^{(2)}_{1}}}} + 2 \operatorname{{{B^{(2)}_{0}}}} m_{n_i}^{2}\right)\Delta_{ii}^{ab}.
 \end{array}
\end{equation}

The diagrams 2 ($W^{\pm} n_i n_j$), 3 ($n_i W^- W^+$), 6 ($n_i G^- G^+$), are finite because the form factor only contain three point Passarino-Veltman functions. But, diagrams 4 ($n_i W^- G^+$) and 5 ($n_i G^- W^+$) contain the $\operatorname{B}_0^{(12)}$ function which is divergent. However, this divergence is canceled by the GIM mechanism. Analogously, it occurs to the divergences associated to diagrams 7 ($n_i W^{\pm}$) and 9 ($W^{\pm} n_i$). Finally, diagrams 1 ($G^{\pm} n_i n_j$), 8 ($n_i W^{\pm}$), 10 ($W^{\pm} n_i$) contain divergences, which are canceled by adding them together~\cite{PhysRevD.71.035011,PhysRevD.91.015001,THAO2017159}. This is shown in some detail in Appendix~\ref{Appendix:Cancellation_divergences}. Also, these results agree with those given in~\cite{PhysRevD.91.015001,Marcano}, taken into account the definitions in~\eqref{ec:B_PV-Thao-Arganda} and $C_0 \to C_0$, $C_1 \to C_{12} - C_{11}$ and $C_2 \to C_{12}$.

Finally, the total form factor $A_{R,L}^{h}(\text{total})_{ab}$ is the result of add the 10 diagrams involved in the model and summing over the neutrino generations, then
	\begin{equation}\label{ec:ALR_total_seesaw}
	A_{R,L}^{h}(\text{total})_{ab} = \sum_{i,j=1}^{6} A_{R,L}^h(n_i n_j)_{ab} +  \sum_{i=1}^{6} A_{R,L}^h(n_i)_{ab},
	\end{equation}
	where
	\begin{subequations}\label{ec:ARLninj}
		\begin{align}
		A_{R,L}^h(n_i n_j)_{ab} &= A_{R,L}^h(G^{\pm} n_i n_j)_{ab} + A_{R,L}^h(W^{\pm} n_i n_j)_{ab},\\\notag
		A_{R,L}^h(n_i)_{ab} &= A_{R,L}^h(n_i G^- G^+)_{ab} + A_{R,L}^h(n_i W^- G^+)_{ab} + A_{R,L}^h(n_i G^- W^+)_{ab} + A_{R,L}^h(n_i W^- W^+)_{ab} \\ 
		& \quad +  
		A_{R,L}^h(n_i W^{\pm})_{ab} + A_{R,L}^h(n_i G^{\pm})_{ab} + A_{R,L}^h(W^{\pm} n_i)_{ab} + A_{R,L}^h(G^{\pm}  n_i)_{ab}.
		\end{align}
	\end{subequations}

\subsection{Neutrino oscillations data and numerical results for $\bf(h\to l_a l_b)$ }

\subsubsection{Neutrino oscillations data}
Before applying our results for LFV Higgs decays, we recollect the relevant neutrino oscillations data, which are described by the mixing matrix $\mathbf{U}_{\mathrm{PMNS}}$, in the standard parametrization is given by 
$\mathbf{U}_{\mathrm{PMNS}} = \mathbf{U} \mathbf{P}_M$, where 
\begin{equation}
\mathbf{U}=\left(\begin{array}{ccc}
c_{12} c_{13} & s_{12} c_{13} & s_{13} e^{-i \delta_{CP}} \\
-s_{12} c_{23}-c_{12} s_{23} s_{13} e^{i \delta_{CP}} & c_{12} c_{23}-s_{12} s_{13} s_{23} e^{i \delta_{CP}} & c_{13} s_{23} \\
s_{12} s_{23}-c_{12} s_{13} c_{23} e^{i \delta_{CP}} & -c_{12} s_{23}-s_{12} s_{13} c_{23} e^{i \delta_{CP}} & c_{13} c_{23}
\end{array}\right),
\end{equation}
with $c_{ij} = \cos{\theta_{ij}}$,$s_{ij} = \sin{\theta_{ij}}$ and $\delta_{CP}$ is the Dirac CP phase. $\mathbf{P}_M$ is a diagonal matrix which appears in the case of Majorana neutrinos, expressed as
$\mathbf{P}_M = \text{diag}(1,e^{i\delta_\alpha}, e^{i\delta_\beta})$, 
and $\delta_\alpha$, $\delta_\beta$ are Majorana phases which do not appear in the $neutrino \to neutrino$ oscillations experiments. 

Experimental values for mixing angles $\theta_{12}$, $\theta_{13}$, 
$\theta_{23}$, two squared-mass splittings $\Delta m_{ij}$ and Dirac-like $\delta_{CP}$ violation phase are known, for the Normal Order (NO) or the Inverted Order (IO) of neutrino masses. For the following discussion  NO ($m_1 < m_2 < m_3$), so, from the global fit ~\cite{Esteban:2018azc} we extract the values of neutrino oscillations parameters presented in Table~\ref{tab:oscilaciones}.
{\renewcommand{\arraystretch}{1.4}
	\begin{table}[h!]
		\begin{center}
			\begin{tabular}{||c|c|c||}
				\hline 
				& BFP $\pm 1 \sigma$   & $3 \sigma$ \\ 
				\hline \hline
				$\sin^2{(\theta_{12})}$& $0.310^{+0.013}_{-0.012}$ & $0.275-0.350$  \\ 
				\hline 
				$\sin^2{(\theta_{23})}$& $0.582^{+0.015}_{-0.019}$  & $0.428 - 0.624$  \\ 
				\hline 
				$\sin^2{(\theta_{13})}$& $0.02240^{+0.00065}_{-0.00066}$ & $0.02244 - 0.02437$ \\ 
				\hline 
				$\frac{\Delta m_{21}^2}{10^{-5} \text{eV}^2}$& $7.39  ^{+0.021}_{-0.020}$ & $6.79 - 8.01$  \\ 
				\hline 
				$\frac{\Delta m_{31}^2}{10^{-5} \text{eV}^2}$&$2.525  ^{+0.033}_{-0.031}$  & $2.431 - 2.622$ \\ 
				\hline 
				$\delta_{CP}/\text{\textdegree}$&$217^{+40}_{-18}$  & $135-366$ \\ 
				\hline 
			\end{tabular}
		\end{center}
		\caption{Neutrino oscillation data for Normal Order. The second (third) column shows the Best Fit Point (BFP) in $1\sigma$ ($3\sigma$) range~\cite{Esteban:2018azc}.}
		\label{tab:oscilaciones}
\end{table}}
However, for practical reasons, we set the values of the mixing angles to their BFP values $\theta^{\text{BFP}}_{12}$, $\theta^{\text{BFP}}_{13}$, $\theta^{\text{BFP}}_{23}$, and the Dirac and Majorana phases to zero, $\delta_{CP} = \delta_\alpha = \delta_\beta = 0$. As a consequence, the numerical mixing matrix $\mathbf{U}_{\mathrm{PMNS}}$ in the standard parametrization  is given by
\begin{equation}\label{ec:Upmns}
\mathbf{U}_{\mathrm{PMNS}} \approx \left(\begin{matrix}
0.8213 & 0.5505 & 0.1497\cr
-0.4630 & 0.4900 & 0.7386\cr
0.3332 & -0.6759& 0.6573
\end{matrix}\right).
\end{equation}
On the other hand, considering a Normal Order (NO) for neutrinos, we can rewrite the neutrino mass of $\nu_{2,3}$ in terms of $m_1$ and squared-mass splittings as follows
\begin{equation}\label{ec:light_mnu}
m_i = \sqrt{m_1^2 + \Delta m_{i1}^2 },\quad i=2,3.
\end{equation}
Planck satellite imposes an upper bound on the sum of the neutrino masses~\cite{Aghanim:2018eyx},
\begin{equation}\label{ec:Planck_bound}
\sum_{i=1}^{3} m_i < 0.12\,\text{eV},
\end{equation} 
the maximum value of $m_1$ to fulfill the before mentioned bound is $m_1^{\text{max}}=3.01 \times 10^{-11}$ GeV. 

\subsubsection{Numerical analysis for $\br(h \to l_a l_b)$}
Light neutrino masses can be explained naturally by means of See-Saw mechanism which imposes a large separation between the scales of the light and heavy neutrinos, $|\mathbf{M}_D|<<|\mathbf{M}_N|$, so, we have that
\begin{equation}\label{ec:seesaw}
	\hat{\mathbf{m}}_\nu \approx \mathbf{M}_D \mathbf{M}_N^{-1} \mathbf{M}_D^T.
\end{equation}
In turn, Dirac neutrino mass matrix can be parameterized in the See-Saw mechanism in terms of neutrino masses, PMNS matrix and an orthogonal matrix  $\mathbf{\xi}$, $3 \times 3$, as follows~\cite{CASAS2001171}:
\begin{equation}\label{ec:seesawMD}
\mathbf{M}_{D}^{\top}=i \mathbf{U}_{N}^{*}\left(\hat{\mathbf{M}}_{N}\right)^{1 / 2} \mathbf{\xi}\left(\hat{\mathbf{m}}_{\nu}\right)^{1 / 2} \mathbf{U}_{\mathrm{PMNS}}^{\dagger},
\end{equation}
where $\hat{\mathbf{m}}_\nu = \text{diag}(m_{n_1},m_{n_2},m_{n_3})$, $\hat{\mathbf{M}}_N = \text{diag}(m_{n_4},m_{n_5},m_{n_6})$ and $\mathbf{U}_{N}$ is an unitary matrix which diagonalizes $\mathbf{M}_{N}$. 
\begin{figure}
	\centering
	\includegraphics[width=0.7\linewidth]{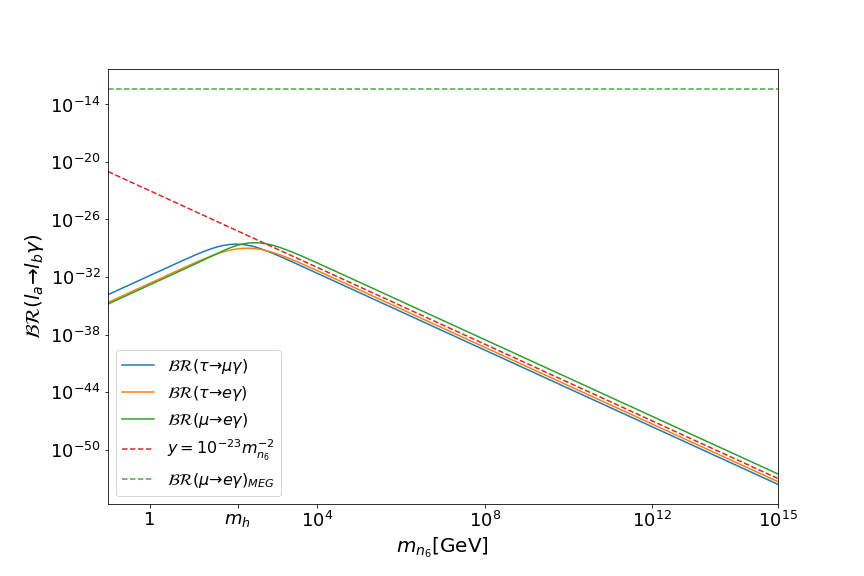}
	\caption{The plot shows $\mathcal{BR}(l_a \to l_b \gamma)$ (solid lines) in the case of  degenerate heavy neutrinos ($m_{n_4} = m_{n_5} = m_{n_6}$). The green dashed lines corresponds to upper bound of $\mu \to e \gamma$ whereas the red dashed line show the curve $y = 10^{-23} m_{n_6}^{-2}$.}
	\label{fig:brlalbgamma}
\end{figure}
\begin{figure}
	\centering
	\begin{subfigure}[t]{.7\linewidth}
		\centering\includegraphics[width=\linewidth]{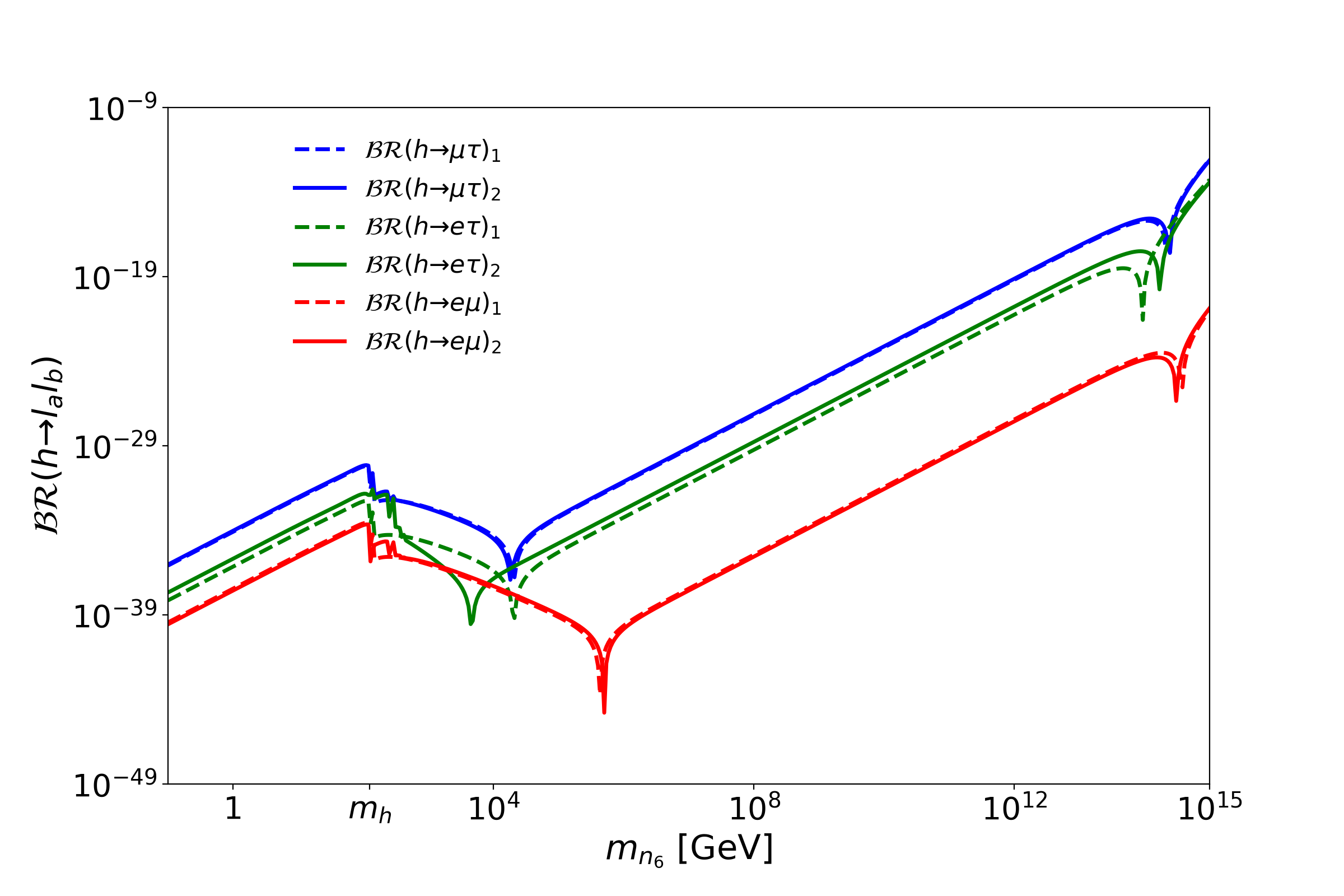}
		\label{fig:BR_comparison}
	\end{subfigure}
	\caption{Numerical results for $\br(h \rightarrow l_a l_b)$ in the $\nu SM$ for the two cases: $(i)$ degenerated heavy spectrum (solid lines) with $m_{n_4} = m_{n_5} = m_{n_6}$; $(ii)$ non-degenerated heavy spectrum (dashed lines) with $m_{n_4} = m_{n_6}/3$ and $m_{n_5} = m_{n_6}/2$.}
	\label{fig:BR-seesaw}
\end{figure}

For the numerical implementation of the LFV Higgs decays, we consider 
$\mathbf{U}_{N} =\mathbf{\xi} = \mathbf{I}$, with $\mathbf{I}$ being the identity matrix, as a consequence $\mathbf{M}_N = \hat{\mathbf{M}}_N$, and we diagonalize numerically the mass matrix~\eqref{ec:seesaw-mass}. 
The mass matrix depends on the neutrino masses and the mixing matrix $\mathbf{U}_{\mathrm{PMNS}}$. However, it follows from  equation~\eqref{ec:light_mnu} valid for the three light neutrino masses, that only the lightest one is independent. Thus, we assume $m_{n_1} = 10^{-12}$ GeV, which is allowed by the Planck upper bound~\eqref{ec:Planck_bound}. The heavy neutrino masses will be the free parameters in this work. We consider two cases, $(i)$ degenerate case with $m_{n_4} = m_{n_5} = m_{n_6}$, and $(ii)$ non-degenerate case with $m_{n_4} = m_{n_6}/3$ and $m_{n_5} = m_{n_6}/2$. For both cases, we consider $m_{n_6}$ in the range $(10^{-1} \text{ GeV}, 10^{15}\text{ GeV})$.  This range respects the two conditions: above this scale it meets the perturvativity constraint~\cite{PhysRevD.91.015001}, and below it meets the See-Saw condition $|\mathbf{M}_D|<<|\mathbf{M}_N|$.
Here we remark one difference in this work compared with~\cite{THAO2017159}, namely we use mpmath functions to diagonalize numerically the mass matrix $\mathbf{M}^\nu$~\eqref{ec:seesaw-mass}, which allows to maintain a numerical stable behavior.

In addition, for the degenerate case, from;~\eqref{ec:seesawMD}, we can deduce that 
\begin{align}\label{ec:MD_CI_seesaw}
\mathbf{M}_D &= i \sqrt{m_{n_6}} \mathbf{U}^{*}_{\mathrm{PMNS}} \left(\hat{\mathbf{m}}_{\nu}\right)^{1 / 2} \quad \Rightarrow \quad \frac{v}{\sqrt{m_{n_6}}} \mathbf{Y_\nu} = i \mathbf{U}^{*}_{\mathrm{PMNS}} \left(\hat{\mathbf{m}}_{\nu}\right)^{1 / 2}\\\label{ec:YY_CI_seesaw}
\frac{v^2 }{ m_{n_6}} (\mathbf{Y_\nu Y_\nu^{\dagger}})_{ab} &=  (\mathbf{U}^{*}_{\mathrm{PMNS}} \hat{\mathbf{m}}_{\nu} \mathbf{U}^{\top}_{\mathrm{PMNS}})_{ab} \quad \Rightarrow \quad  (\mathbf{Y_\nu Y_\nu^{\dagger}})_{ab} \propto  m_{n_6}.
\end{align}
On the other hand, from equation~\eqref{ec:seesaw-indentities} we have
\begin{align}\notag
(\mathbf{M}_D)_{ab} &= (\mathbf{U}^{\nu *} \hat{\mathbf{M}}^{\nu} \mathbf{U}^{\nu\dagger})_{a(b+3)}\\\notag
&= \sum_{k = 1}^6 U^{\nu *}_{ak} m_{n_k} U^{\nu *}_{(b+3)k}\\\label{ec:MD_UU_ss}
&\approx  m_{n_6} \sum_{k = 4}^6 U^{\nu *}_{ak} U^{\nu *}_{(b+3)k} \qquad \textrm{with} \, m_{n_{1,2,3}} << m_{n_6}.
\end{align}
On the other hand, from~\eqref{ec:MD_CI_seesaw} and~\eqref{ec:MD_UU_ss}, we can deduce
\begin{equation}\label{ec:UUab_SS}
\sum_{k = 4}^6 U^{\nu *}_{ak} U^{\nu *}_{(b+3)k} \approx \frac{i}{\sqrt{m_{n_6}}} (\mathbf{U}^{*}_{\mathrm{PMNS}} \left(\hat{\mathbf{m}}_{\nu}\right)^{1 / 2})_{ab}.
\end{equation}

Figure~\ref{fig:brlalbgamma} shows the behavior of $\mathcal{BR}(l_a \to l_b \gamma)$ in the degenerate scenario ($m_{n_4} = m_{n_5} = m_{n_6}$). On the one hand, for $m_{n_6} > 10^3$ GeV, the curve $y = 10^{-23} m_{n_6}^{-2}$ is a good approximation to $\mathcal{BR}(l_a \to l_b \gamma)$. On the other hand, an approximation to $\mathcal{BR}(l_a \to l_b \gamma)$ was previously found by~\cite{PhysRevD.91.015001}, which is valid in the regime of heavy neutrino masses, viz, $\mathcal{BR}(l_a \to l_b \gamma)_{\text{approx}} \propto |\frac{v^2}{2 m_{n_6}^2} (\mathbf{Y_{\nu} Y_\nu^{\dagger}})_{ab}|^2$. So, from equation~\eqref{ec:YY_CI_seesaw} we conclude that $\mathcal{BR}(l_a \to l_b \gamma)_{\text{approx}} \propto m_{n_6}^{-2}$, and that the experimental upper bound from MEG (see Table~\ref{tb:bounds-lfv}) to $\mu \to e \gamma$ does not constraint the heavy neutrino mass $m_{n_6}$, as illustrated in Figure~\ref{fig:brlalbgamma}.

Figure~\ref{fig:BR-seesaw} shows $\mathcal{BR}(h  \to l_a l_b)$ as function of $m_{n_6}$, as we expected, $\br(h \to \mu \tau)$ is the largest branching ratio in agreement with literature~\cite{PILAFTSIS199268,PhysRevD.47.1080,PhysRevD.71.035011,THAO2017159}. The highest value is $\br(h \to \mu \tau) \approx 10^{-12}$ with $m_{n_6} \simeq 10^{15} \, \textrm{GeV}$ for degenerate and non-degenerate cases. 
Now, considering $\br(l_a \to l_b \gamma)$ in the type I See-Saw in Figure~\ref{fig:brlalbgamma}, the behavior in the region of $m_{n_6} \in [10^{-1},10^4]$ is similar to $\br(h \to l_a l_b)$. However, in this region, for the LFV Higgs decays the interaction $h n_i n_j$ is not dominant and diagrams with only one neutrino into the loop are important. Further, when $m_{n_6} > 10^{4}$ GeV the interaction $h n_i n_j$ does contribute to loop amplitude for $h \to l_a l_b$, but it does not contribute to $l_a \to l_b \gamma$, as a consequence we obtain large $\br(h \to l_a l_b)$.

\begin{figure}
	\centering
		\includegraphics[width=.7\linewidth]{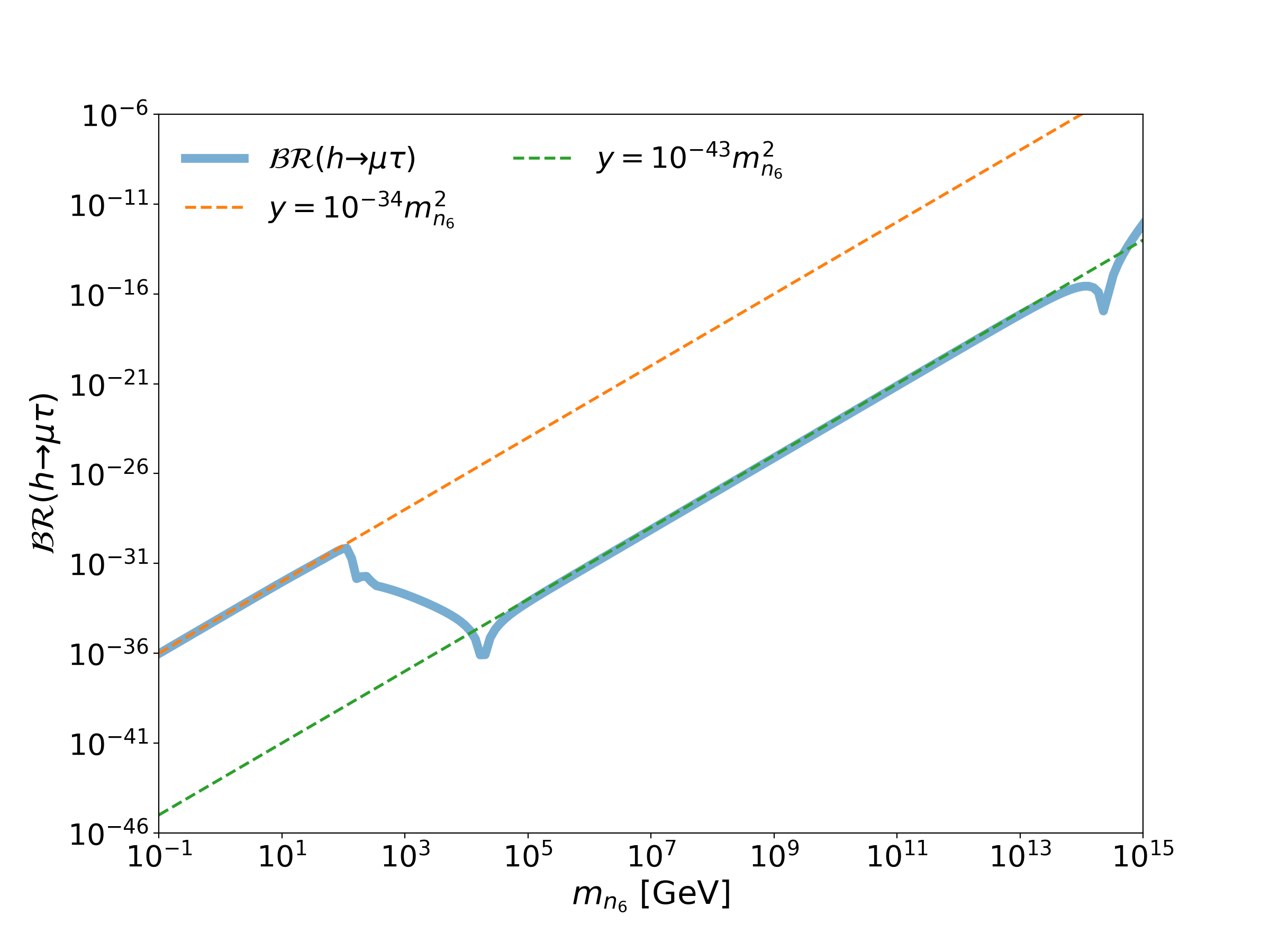}
	\caption{Plot shows the behavior of $\br(h \to \mu \tau)$ vs $m_{n_6}$ for the non-degenerate case, and the dashed lines $y \propto m_{n_6}^2$ fit the $\br(h \to \mu \tau)$ in different regions.
}
	\label{fig:BRhl2l3_pendientes}
\end{figure}

On one hand, in Figure~\ref{fig:BRhl2l3_pendientes}, we show the total branching ratio, and the line $y = 10^{-34} m_{n_6}^2$, which coincides with $\br(h \to \mu \tau)$ in the region $m_{n_6} \in [10^{-1}\text{ GeV},m_h]$, while the line $y = 10^{-43} m_{n_6}^2$ fits the $\br(h \to \mu \tau)$ in $m_{n_6} \in (10^{4},10^{15}]$. On the other hand, in~\cite{PILAFTSIS199268,PhysRevD.47.1080}, an approximate result to $\br(h \to l_a l_b) \propto m_{n}^4 |F_N|^2$ with $m_n$ denotes the heavy neutrino mass and $F_N = U_{bj}^{\nu} C_{ij} U_{ai}^{\nu *}$ is obtained. In our case, from equation~\eqref{ec:UUab_SS}, this approximate result is equivalent to  $\br(h \to l_a l_b) \propto m_{n_6}^2$ with $F_N  \propto m_{n}^{-1}$, which is also in agreement with the reported in~\cite{THAO2017159}. However, we found a slight difference in the $\br(h \to \mu \tau)$ and $\br(h \to e \tau)$ in contrast to Thao et al.~\cite{THAO2017159}, specifically showing a gap between these branching ratios, which is a result of the high precision used in the numerical analysis.

As a result of the proportionality of Yukawa couplings to $\sqrt{m_{n_6}}$, induced by the  Casas-Ibarra parametrization (see equation~\eqref{ec:MD_CI_seesaw}), an apparent non-decoupling behavior of $\br(h \to \mu \tau)$ is presented, which is also analyzed in~\cite{PhysRevD.91.015001} in the context of the Inverse See-Saw (ISS) model. Although in this case, the Casas-Ibarra parametrization induces that the Yukawa couplings increase as $M_R/\sqrt{\mu_X}$, where $M_R$ is the mass of right-handed heavy neutrinos and $\mu_X$ is the mass of the three additional singlets in the ISS, both in the degenerate case. As a result, the maximum values of $\br(h \to \mu \tau)_{ISS}$ are bigger than our results, but the perturvativity limits on Yukawa couplings constraint  the heavy neutrino mass $M_R$ stronger than in the type I See-Saw model.

In case of~\cite{PhysRevD.102.113006}, just two new heavy right-handed neutrinos are included with a particular mixing to left-handed neutrinos, as a consequence  one simple form of neutrino mass matrix is induced, and the LFV Higgs decays are proportional to the light-heavy mixing angle $s_{\nu_i}$. In this work they consider Yukawa couplings proportional to $y_i \propto \sqrt{M_{n_1}M_{n_2}}s_{\nu_i}$, which is greater than our assumption with the Casas-Ibarra parametrization. Hence, the perturvativity of Yukawa couplings  forbids a large heavy mass scale.

Finally, a descendant behavior of $\br(h \to l_a l_b)$ is shown in~\cite{PhysRevD.71.035011}, which agrees with the behavior of $\br(h \to \mu \tau, e \tau)$ and $\br(h \to e \mu)$, in the mass regions $m_h < m_{n_6} \lesssim 10^{4}$ and $m_h < m_{n_6} \lesssim 10^{6}$ respectively, as is shown in the Figures~\ref{fig:BR-seesaw} and~\ref{fig:BRhl2l3_pendientes}. 

As a  conclusion, the largest values of $\br(h \to \mu \tau)$ are found to be of the order of $\mathcal{O}(10^{-12})$ for $m_{n_6} \approx 10^{15}$ GeV. These results are well below the current experimental bounds from LHC~\cite{2020135069}, and it probably works as the lowest possible value for $\br(h \to \tau \mu)$. Furthermore, we notice that $\br(h \to \tau e)$ and $\br(h \to \mu e)$ have lower values than $\br(h \to \tau \mu)$ for most regions of the parameter space.

\section{LFV Higgs Decays within the Scotogenic Model\label{sec:lfvhiggsscot}}
As a second application of our formalism, we shall study the LFV Higgs decays within the Scotogenic Model~\cite{PhysRevD.73.077301}. The LFV Higgs decays in this model were studied previously. In ~\cite{Herrero-Garcia2016} an approximate upper bound was founded, and ~\cite{Hundi:2022iva} studied this signals in a region of the parameter space where maximum values to LFV Higgs decays are reached around of $\br(h \to \mu \tau)_{\text{max}} \approx 10^{-7}$. We present the features of this model in the following.

\subsection{Model content and parameters}
The matter content of the Scotogenic Model with the group $SU(2)_L\otimes U(1)_Y\otimes Z_2$ is given by~\cite{PhysRevD.73.077301}: 
\[
\begin{array}{c}
\left(\nu_{L a}, l_{L a}\right) \sim(2,-1 / 2,+),\quad l_{R a} \sim(1,1,+), \quad N_{R k} \sim(1,0,-), \cr
\left(\phi^{+}, \phi^{0}\right) \sim(2,1 / 2,+), \quad\left(\eta^{+}, \eta^{0}\right) \sim(2,1 / 2,-),
\end{array}
\]
where an inert doublet $\eta$ and three right-handed neutrinos $N_{Ri}$ odd under $Z_2$ symmetry have been included. Yukawa sector allows the interaction between left and right-handed neutrinos with scalar doublet $\eta$:
\begin{equation}
\mathcal{L}_{Y}=f_{a b}\left(\phi^{-} \nu_{L a}+\overline{\phi_{0}} l_{L a}\right) l_{R b}+ Y'_{a b}\left(\nu_{L a} \eta_{0}-l_{L a} \eta^{+}\right) N_{R b}+ \textrm{H. c.}
\end{equation}
and the Majorana term for right-handed neutrinos is added,
\begin{equation}
\frac{1}{2} M_{k} \overline{N_{R k}}^c N_{R k}+\textrm{H. c.},
\end{equation}
being $M_k$ the mass of $N_{R k}$. In this case, the more general scalar potential under $Z_2$ symmetry is as follows,
\begin{equation}\label{ec:potencial_escalar}
\begin{aligned} V(\Phi, \eta)&= \mu_{1}^{2}\left(\Phi^{\dagger} \Phi\right)+\mu_{2}^{2}\left(\eta^{\dagger} \eta\right)+\frac{\lambda_{1}}{2}\left(\Phi^{\dagger} \Phi\right)^{2}+\frac{\lambda_{2}}{2}\left(\eta^{\dagger} \eta\right)^{2} \\ &+\lambda_{3}\left(\Phi^{\dagger} \Phi\right)\left(\eta^{\dagger} \eta\right)+\lambda_{4}\left(\Phi^{\dagger} \eta\right)\left(\eta^{\dagger} \Phi\right)+\frac{\lambda_{5}}{2}\left[\left(\Phi^{\dagger} \eta\right)^{2}+ \textrm{H. c.} \right]. \end{aligned}
\end{equation}
The mass spectrum to new scalar particles is
\begin{equation}
\label{ec:masas-escalares}
\begin{aligned} m^{2}\left(\sqrt{2} R e \phi_{0}\right)&=&m_h^2 &=2 \lambda_{1} v^{2}, \\ m^{2}\left(\eta^{ \pm}\right)&=&m_{\eta}^2 &=\frac{\lambda_{3} v^{2}}{2} + \mu_{2}^{2}, \\ 
m^{2}\left(\sqrt{2} \mathit{Re} \eta_{0}\right)&=&m_{R}^2 &=2 \mu_{2}^{2}+v^{2}\left(\lambda_{3}+\lambda_{4}+\lambda_{5}\right), \\ 
m^{2}\left(\sqrt{2} \mathit{Im} \eta_{0}\right)&=&m_I^2 &=2 \mu_{2}^{2}+v^{2}\left(\lambda_{3}+\lambda_{4}-\lambda_{5}\right),
\end{aligned}
\end{equation}
and SM-like Higgs is associated to $h = \sqrt{2} \mathit{Re} \phi_0$.

Neutrino mass is generated by a radiative See-Saw mechanism and is given by
\begin{equation}\label{ec:Mnu}
\left(\mathbf{\mathcal{M}}_\nu\right)_{i j}= Y'_{i k} Y'_{j k} M_{k} \frac{1}{16 \pi^{2}}\left[\left(\frac{m_{R}^{2}}{M_{k}^{2}-m_{R}^{2}} \log \frac{m_{R}^{2}}{M_{k}^{2}}\right)-\left(\frac{m_{I}^{2}}{M_{k}^{2}-m_{I}^{2}} \log \frac{m_{I}^{2}}{M_{k}^{2}}\right)\right],
\end{equation}
where $m_R$, $m_I$, $m_\eta$ and $M_k$ are the masses of $\eta_R$, $\eta_I$, $\eta^{\pm}$ and $N_{Ri}$ respectively. Also, $k$ index runs over heavy neutrino masses.
This matrix can be rewritten in a matrix notation as 
$\mathbf{\mathcal{M}}_\nu = \mathbf{Y'}\mathbf{D}_{\Lambda}\mathbf{Y'}^{\top}$,
where
\begin{equation}\label{ec:Lambda_k}
	\begin{array}{rcl}
\mathbf{D}_{\Lambda} &= &\text{diag}(\Lambda_1,\Lambda_2,\Lambda_3),\\
\Lambda_k &= & \frac{M_{k}}{16 \pi^{2}}\left[\left(\frac{m_{R}^{2}}{M_{k}^{2}-m_{R}^{2}} \log \frac{m_{R}^{2}}{M_{k}^{2}}\right)-\left(\frac{m_{I}^{2}}{M_{k}^{2}-m_{I}^{2}} \log \frac{m_{I}^{2}}{M_{k}^{2}}\right)\right].
\end{array}
\end{equation}
Then, if we choose the basis where Yukawa matrix $\mathbf{Y}'$ is diagonal, $\mathbf{\mathcal{M}}_\nu$ is diagonal too. So, we can rewrite the Yukawa matrix as follows,
\begin{equation}\label{ec:Y'}
    \mathbf{Y'} = \text{diag}\left(\sqrt{\frac{m_{\nu_1}}{\Lambda_1}},\sqrt{\frac{m_{\nu_2}}{\Lambda_2}},\sqrt{\frac{m_{\nu_3}}{\Lambda_3}}\right) \quad \text{or} \quad  Y'_{ij} = \sqrt{\frac{m_{\nu_i}}{\Lambda_i}} \delta_{ij},
\end{equation}
where $m_{\nu_i}$ ($i,j=1,2,3$) are the light neutrino masses in normal order.
However, as we have noticed, in this basis the matrix to diagonalize the neutrino mass  matrix, $\mathbf{V}_L^{\nu}$ is equal to identity. On the other hand, the mass matrix of charged leptons will be rotated by $\mathbf{V}_L^{l}$ to go to the physical basis. Thus, the neutrino mixing matrix is $\mathbf{U}_{\mathrm{PMNS}} = \mathbf{V}_L^{\nu \dagger}\mathbf{V}_L^{l} = \mathbf{V}_L^{l}$. Finally, in the physical basis, new Yukawa couplings for the interaction between  left-handed lepton doublet and singlet neutrinos given by
\begin{equation}\label{ec:scotoYuk}
    \mathbf{Y} = \mathbf{Y'}\mathbf{U}_{\mathrm{PMNS}}.
\end{equation}
However, by means of equation~\eqref{ec:Y'} and~\eqref{ec:scotoYuk} we have that
\begin{equation}\label{ec:YdagaY}
    ( \mathbf{Y}^{\dagger} \mathbf{Y})_{ij} = Y_{ki}^* Y_{kj} = \frac{m_{\nu_k}}{\Lambda_k}U_{ki}^* U_{kj},
\end{equation}
which will allow the use of the GIM mechanism in the loop calculation.
\begin{table}
	\begin{center}
		\begin{tabular}{||c|c||c|c||}
			\hline \hline
			\multicolumn{1}{|c|}{\bf Vertex} & {\bf Coupling} & \multicolumn{1}{|c|}{\bf Vertex} & {\bf Coupling} \\ \hline
			$h G^{+} G^{-}$ & $-i \frac{m_{h}^{2}}{v}$ &
			$h \eta^{+} \eta^{-}$ & $ -\frac{2i\left(m_{\eta}^{2}-\mu_{2}^{2}\right)}{v} $ \\ \hline
			$G^{-}l^{+}_a \nu_{k}$ & $i \frac{\sqrt{2}}{v} m_a U^{*}_{ka} P_{L}$&
			$G^{+}l^{-}_b \nu_{k}$ & $i \frac{\sqrt{2}}{v} m_b U_{kb} P_{R}$ \\ \hline
			$h G^{+} W^{-}$ & $-\frac{i}{2}g (p_{+} - p_{0})_\mu$ &
			$h  W^{+} G^{-}$ & $-\frac{i}{2}g (p_{0} - p_{-})_\mu$ \\ \hline
			$W^{-} l^{+}_a \nu_{k}$ & $-i \frac{g}{\sqrt{2}}U^{*}_{ka} \gamma_\mu P_L$ &
			$W^{+} l^{-}_b \nu_{k}$ & $-i \frac{g}{\sqrt{2}}U_{kb} \gamma_\mu P_L$ \\ \hline
			$\eta^{-} l^{+}_a N_{k}$ & $i Y^{*}_{ka} P_R$ &
			$\eta^{+} l^{-}_a N_{k}$ & $i Y_{kb} P_L$  \\ \hline
			$h W^{+}_\mu W^{-}_\nu$& $ig m_W g_{\mu \nu}$ & & \\ \hline \hline
		\end{tabular}
	\end{center}
	\caption{Couplings involved in the LFV Higgs decays for the Scotogenic model. The $p_0$, $p_+$ and $p_-$ are incoming momenta of $h$, $G^+$ and $G^-$, respectively.}
	\label{tab:Vertices_Escoto}
\end{table}
%

\subsection{Model constraints}
Although the Scotogenic model is a minimal extension of SM, it describes a wide spectrum of phenomena such as: massive neutrinos, fermionic dark matter or LFV processes, each of these phenomena gives some constraints
on the parameter space. Next, we describe the constraints that are used to derive the allowed regions of parameter space. 
\begin{itemize}
\item {\bf{Electroweak parameters $(S,T,U)$. }}
For the case  where $U = 0$, the experimental values of the parameters $S$ and $T$ are:  $S=0.0 \pm 0.07, \, T=0.05 \pm 0.06$~\cite{Zyla:2020zbs}. The analytic expressions for these parameters can be obtained from Ref.~\cite{GRIMUS200881}. As a consequence of these constraints, the masses of  scalar inert particles are very close to each other. 
\item {\bf{Charged Lepton Flavor Violation. }}
The LFV processes $l_b \to l_a \gamma$ are a consequence of neutrino masses, and experimentally they have been highly restricted, as it is summarized in Table~\ref{tb:bounds-lfv}.
\begin{table}[H]
	\begin{center}
		\begin{tabular}{|c|c|}
			\hline
			Process & Experimental bound \\
			\hline
			$\mathcal{BR}(\mu \rightarrow e \gamma)$ & $4.2\times 10^{-13}$\cite{TheMEG:2016wtm} \\
			\hline
			$\mathcal{BR}(\tau \rightarrow e \gamma)$ & $3.3\times 10^{-8}$\cite{Aubert:2009ag} \\
			\hline
			$\mathcal{BR}(\tau \rightarrow \mu \gamma)$ & $4.2\times 10^{-8}$\cite{Uno2021}  \\
			\hline
		\end{tabular}
	\end{center}
	\caption{Experimental upper bounds for CLFV processes.}
	\label{tb:bounds-lfv}
\end{table}

In the Scotogenic model, the branching ratio of these processes are given by~\cite{Toma2014}:
\begin{equation}\label{ec:ljmugamma}
\mathcal{BR}(l_b \rightarrow l_a \gamma) = \frac{3 (4 \pi)^3 \alpha}{4 G_F^2}|A_D|^2 \mathcal{BR}(l_b \rightarrow l_a \nu_b \bar{\nu_a}),
\end{equation}
where
$A_D = \sum_{i=1}^{3}\frac{Y_{ki}^{*}Y_{kj}}{2(4 \pi)^2}\frac{1}{m_\eta^2}F\left(\frac{M_k^2}{m_\eta^2}\right)$,
and $F(x) = \frac{1 -6x + 3 x^2 + 2 x^3 - 6 x^2 \log{x}}{6(1-x)^4}$. The experimental limits are $\mathcal{BR}\left(\tau \rightarrow \mu \nu \bar{\nu}\right) = 0.1739\pm0.0004$, $\mathcal{BR}\left(\tau \rightarrow e \nu \bar{\nu}\right) = 0.1782\pm0.0004$, and $\mathcal{BR}\left(\mu \rightarrow e \nu \bar{\nu}\right) \approx 1$~\cite{Zyla:2020zbs}.

\item {\bf{Fermionic dark matter}}  The experimental value of the relic density of dark matter $\Omega \hat{h}^2$ is
   $ \Omega \hat{h}^2 = 0.12 \pm 0.0012$.  In the Scotogenic model the $Z_2$ symmetry allows for both boson and fermion dark matter candidates, we choose the case of the lightest singlet neutrino $N_1$ as cold fermionic dark matter.
The  parameter $\Omega \hat{h}^2$  is related to annihilation cross section $\sigma_{ann}$ by~\cite{PhysRevD.87.095015,KUBO200618,JUNGMAN1996195}:
\begin{equation}\label{ec:Omegah}
\Omega \hat{h}^2 = \frac{1.07\times 10^9 x_f \text{GeV}^{-1}}{ \sqrt{g_{*}}m_{Pl}(\hat{a} + 3(\hat{b}-\hat{a}/4))/x_f },
\end{equation}
where $x_f = \ln [0.955 (\hat{a} + 6\hat{b}/x_f)M_1 m_{Pl}] / \sqrt{g_{*}x_f} $, and
$\hat{h}$ denotes the Hubble parameter, $m_{Pl} = 1.22\times 10^{19}$ GeV is the Planck mass, $g_*$ is the number of relativistic degrees of freedom below the freeze-out temperature $T_f = M_{N_1}/x_f$;  $\hat{a}$ and $\hat{b}$ are defined by the expansion 
$\sigma_{ann}v_{\text{rel}} = \hat{a} + \hat{b} v_{\text{rel}^2} + O(v_{\text{rel}}^4)$, in terms of the relative speed $v_{\text{rel}}$ of the $N_1 \overline{N}_1$ pair in their center-of-mass frame. 
At tree level, the processes $N_1 \overline{N}_1 \rightarrow l_i^+ l_j^-$ and $N_1 \overline{N}_1 \rightarrow \nu_i \overline{\nu}_j$ mediated by $\eta^{\pm}$ and $\eta_{R,I}$ respectively, are the only ones which contribute to $\sigma_{ann}$. 
In the limit  when all final masses neglected, their combined cross section times $v_{\text{rel}}$ 
is expressed as~\cite{PhysRevD.87.095015}
\begin{align}\label{ec:sigvrelN_1}
\begin{split}
\sigma_{ann}v_{rel} 
&= \sum_{i, j=1,2,3} \frac{\left|Y_{i 1} Y_{j 1}\right|^{2} M_{1}^{2} v_{\mathrm{rel}}^{2}}{48 \pi} \left[\frac{M_{1}^{4}+m_{\eta}^{4}}{\left(M_{1}^{2}+m_{\eta}^{2}\right)^{4}}+\frac{M_{1}^{4}+m_{0}^{4}}{\left(M_{1}^{2}+m_{0}^{2}\right)^{4}}\right],
\end{split} 
\end{align}
with $m_0^2 = (m_R^2 + m_I^2)/2$. Although~\eqref{ec:sigvrelN_1} shows the main dependence of $\sigma_{ann} v_{rel}$, the relic density of dark matter provides a strong constraint, together with LFV process. To improve the numerical precision, we include the charged lepton masses in the numerical analysis. 
\end{itemize}

\subsubsection{Allowed parameter space}
The parameter space consists of heavy neutrino masses $M_{k}$, where $M_{1}$ is the mass of the dark matter candidate, also, it depends on inert scalar masses $m_{\eta}$, $m_{R}$, light neutrino mass $m_1$ and finally parameters $\lambda_{2,5}$ and $\mu_2^2$ from scalar potential\footnote{It does not depend on $m_I$ due to $m_I^2 = m_R^2 - 2 \lambda_5 v^2$.}. We have found an allowed parameter space fixing $\lambda_{2}=0.1$ (which only affects the scalar sector) and $m_1 = m_1^{\text{max}}=3.01\times 10^{-11}$ GeV which fulfill the Planck satellite upper bound on light neutrino masses. Then a random scan is performed considering the follows ranges for the parameter space
\begin{eqnarray}\label{ec:init_space}\notag
1\, \text{GeV} < M_1 < 1\,\text{TeV}; \quad
10 \, \text{TeV}< M_{2,3}< 50 \, \text{TeV};\\\notag
100\, \text{GeV} < m_{R,\eta} < 1\, \text{TeV};  \quad\\
10^{-11} < \lambda_5 < 10^{-8}; \quad
|\lambda_{3,4}| < 4\pi; \quad
M_1< m_{R,I,\eta}, M_{2,3}.
\end{eqnarray}

\begin{figure}%
	\raggedright 
	\begin{tabular}{ll}
		\includegraphics[width=8.0cm,height=7cm]{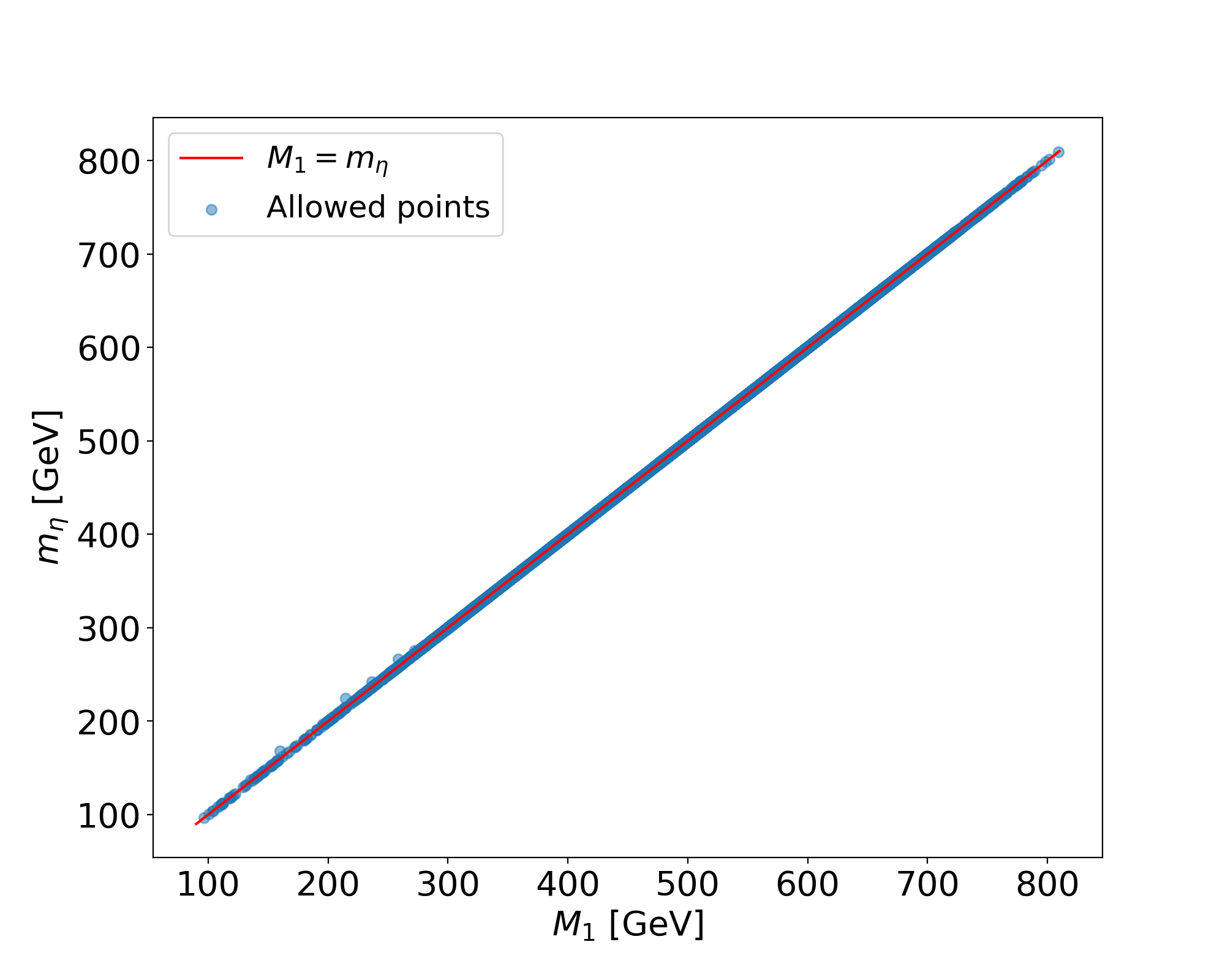}&
		\includegraphics[width=8.0cm,height=7cm]{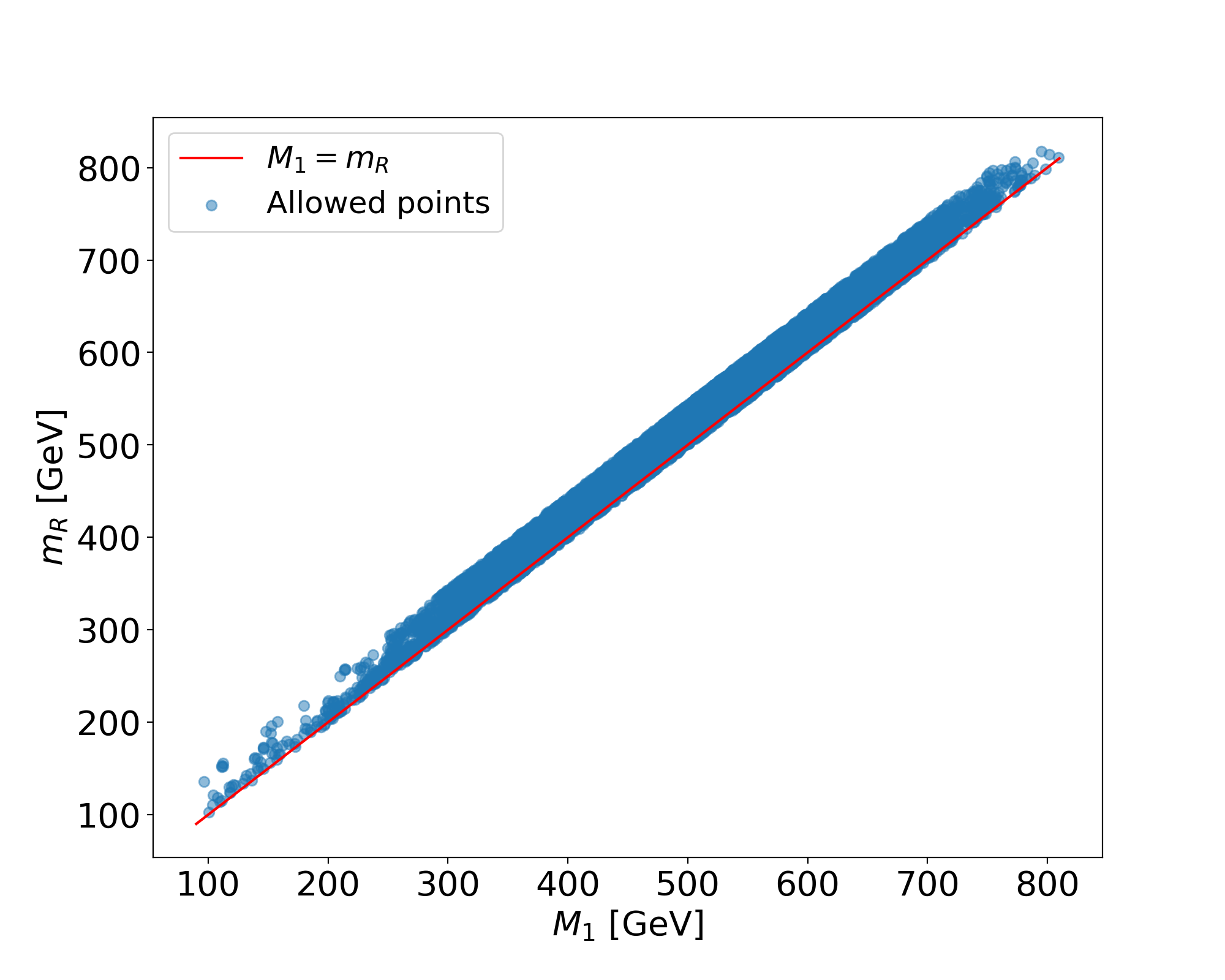} 
	\end{tabular}
	\caption{Allowed parameter space for $M_{1}$, $m_\eta$ and $m_R$. The  left plot shows that $m_\eta \approx M_1$, whereas right plot shows $m_R \gtrapprox M_1$.} 
	\label{fig:mRmetaM1}
\end{figure}

\begin{figure}[t]
	\raggedright
	\begin{tabular}{ll}
		\includegraphics[width=8.0cm,height=7cm]{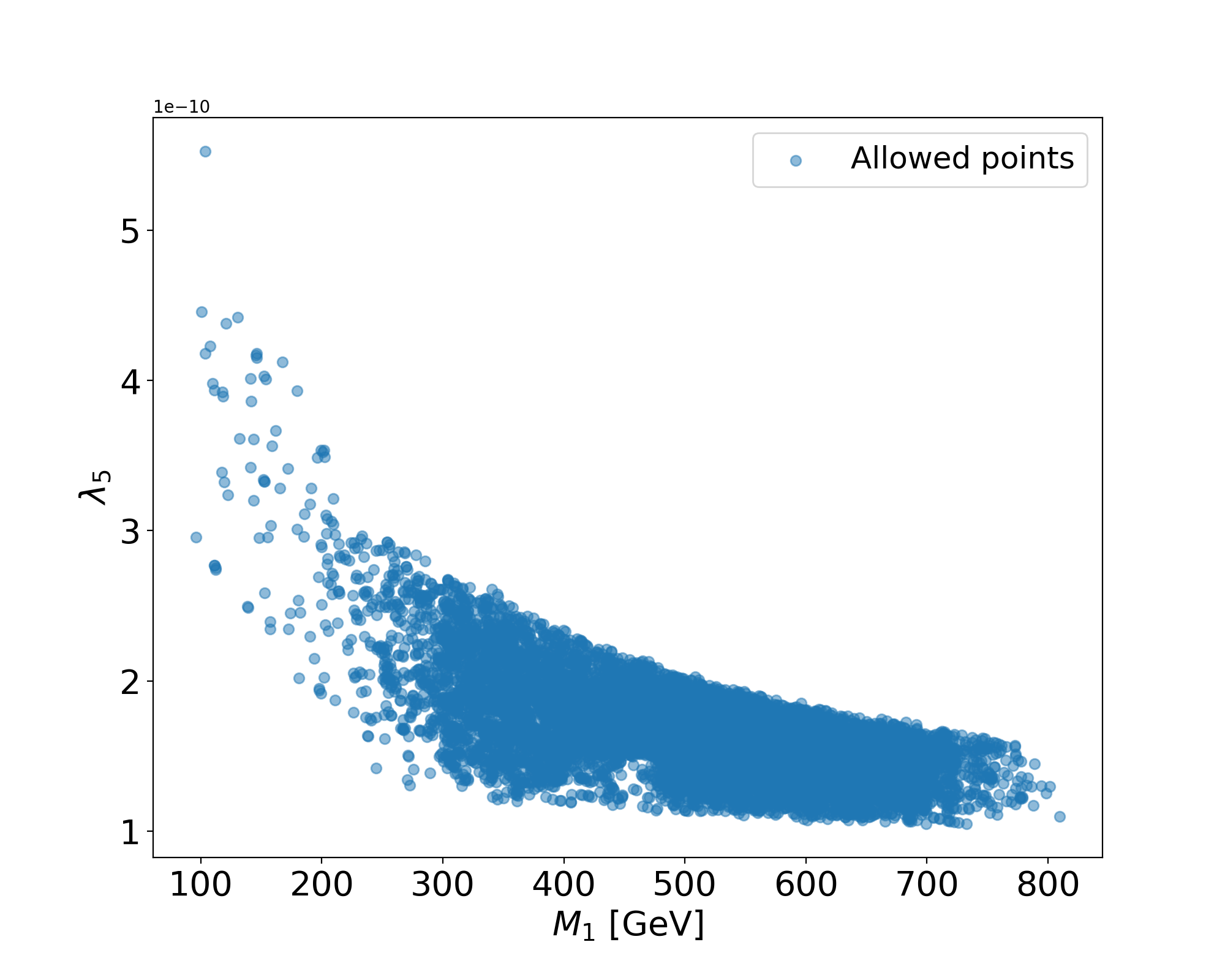}&
		\includegraphics[width=8.0cm,height=6.2cm]{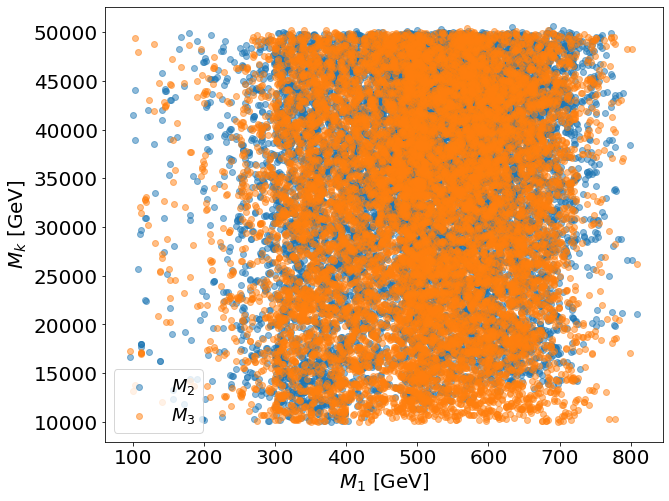} \\
		\multicolumn{2}{c}{\includegraphics[width=8.5cm,height=7cm]{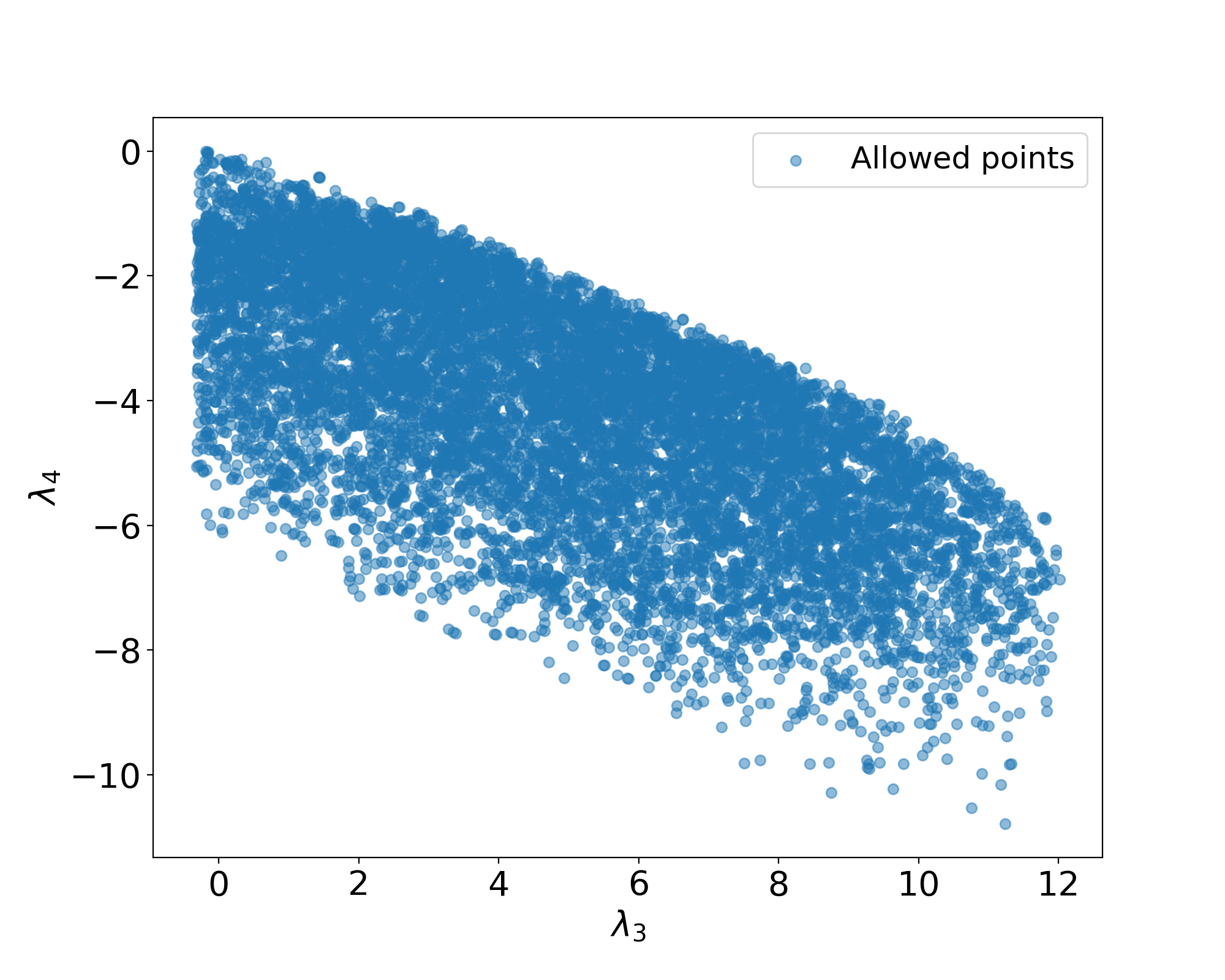}}
	\end{tabular}
	\caption{Allowed parameter space for $M_{1,2,3}$ and  $\lambda_{3,4,5}$. It is observed from plane $\lambda_{5}$ vs $M_1$ that $\lambda_{5} \approx 10^{-10}$. The masses of $M_{2,3} \gg M_1$ comes from top right plot. Finally, we have that $-10.8 < \lambda_{4} \lessapprox 0.02$ and $0 \lessapprox \lambda_{3} \lessapprox 12$ in the bottom plot. } 
	\label{fig:M123l345}
\end{figure}

From the scalar spectrum we can deduce the relation $\mu_2^2 = m_R^2 - m_{\eta}^2 - v^2(\lambda_3/2 + \lambda_4 + \lambda_5)$, which is assumed in this analysis. The final constraint of~\eqref{ec:init_space} guarantees that $N_1$ is a viable candidate for dark matter. The allowed parameter space is projected on different planes as it is shown in Figures~\ref{fig:mRmetaM1} and~\ref{fig:M123l345}. The parameter space is strongly constrained for $m_\eta$, $m_R$ and dark matter mass $M_1$, mainly by the CLFV $l_b \to l_a \gamma$ process and dark matter relic density bounds. In addition, $\lambda_5$ is tiny because of the degeneracy between $m_R$ and $m_I$, which guarantees light neutrino masses and the other constraints. Finally, we also conclude from  this scan that $\lambda_{5} \approx 10^{-10}$ and the heavy neutrino masses $M_{2,3}$ are not constrained in such region.
%

\subsection{Form factors for the amplitude of the decay $h \to l_a^+ l_b^-$}
It is observed as a consequence of $Z_2$ symmetry that diagrams for LFVHD in the Scotogenic model which contain light and heavy neutrinos  are separate, also, in this model, the interaction $h n_i n_j$, with $n_i$ a light or heavy neutrino it is not allowed, as a consequence  we only have diagrams with one neutrino inside the loop. The relevant couplings have been calculated with SARAH package~\cite{STAUB20141773} and are given in Table~\ref{tab:Vertices_Escoto} and the associated diagrams are calculated in the next two subsections.

\subsubsection{Heavy neutrino contributions}
\begin{figure}
  \centering
  \begin{subfigure}[t]{.3\linewidth}
    \centering\includegraphics[width=.8\linewidth]{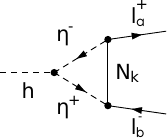}
    \caption{FSS-like.}
    \label{fig:Scoto-N_etau_etad}
  \end{subfigure}
  \begin{subfigure}[t]{.3\linewidth}
    \centering\includegraphics[width=.8\linewidth]{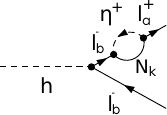}
    \caption{FS-like.}
    \label{fig:Scoto-Neta}
  \end{subfigure}
  \begin{subfigure}[t]{.3\linewidth}
    \centering\includegraphics[width=.8\linewidth]{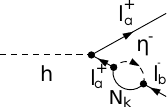}
    \caption{SF-like.}
    \label{fig:Scoto-etaN}
  \end{subfigure}
\caption{Heavy neutrino diagrams of LFV Higgs decays in the Scotogenic model. The sub-captions show the respective generic structure of each diagram.}
\label{fig:Scoto-Ndiagrams}
\end{figure}

In this model, heavy neutrinos only interact with other heavy neutrinos or inert scalar fields. The last ones can interact with SM Higgs boson by means of scalar potential couplings~\eqref{ec:potencial_escalar}. All of these interactions are given in Table~\ref{tab:Vertices_Escoto}. Allowed diagrams are summarized in the first three rows of Table~\ref{tab:scoto_contributions} and shown in Figure~\ref{fig:Scoto-Ndiagrams}. 
\begin{center}
	{
		\begin{table}
			\centering
			\begin{tabular}{||c|c|c|c|c||}
				\hline \hline
				\rule[-1ex]{0pt}{2.5ex} & {\bf Diagram(Figure)} & $P_0$ & $P_1$ & $P_2$ \\
				\hline \hline
				\rule[-1ex]{0pt}{2.5ex} 1& FSS(\ref{fig:Scoto-N_etau_etad})& $N_k$  &$\eta$ &  $\eta$  \\
				\hline
				\rule[-1ex]{0pt}{2.5ex} 2& FS(\ref{fig:Scoto-Neta}) & $N_k$  &$\eta$ &  ---  \\
				\hline
				\rule[-1ex]{0pt}{2.5ex} 3& SF(\ref{fig:Scoto-etaN}) & $N_k$  & --- &  $\eta$ \\
				\hline
				\rule[-1ex]{0pt}{2.5ex} 4& FSS(\ref{fig:Scoto-nGuGd}) & $\nu_k$  &$G^{\pm}$ &  $G^{\mp}$ \\
				\hline
				\rule[-1ex]{0pt}{2.5ex} 5& FSV(\ref{fig:Scoto-nGuWd}) & $\nu_k$ &$G^{\pm}$ & $W^{\mp}$ \\
				\hline
				\rule[-1ex]{0pt}{2.5ex} 6 &FVS(\ref{fig:Scoto-nWuGd}) & $\nu_k$  &$W^{\pm}$ & $G^{\mp}$  \\
				\hline
				\rule[-1ex]{0pt}{2.5ex} 7& FVV(\ref{fig:Scoto-nWuWd}) & $\nu_k$  & $W^{\pm}$ & $W^{\mp}$ \\
				\hline
				\rule[-1ex]{0pt}{2.5ex} 8& FS(\ref{fig:Scoto-nG}) & $\nu_k$  & $G^{\pm}$ &   ---  \\
				\hline
				\rule[-1ex]{0pt}{2.5ex} 9& FV(\ref{fig:Scoto-nW}) & $\nu_k$  & $W^{\pm}$ &   ---  \\
				\hline
				\rule[-1ex]{0pt}{2.5ex} 10& SF(\ref{fig:Scoto-Gn}) & $\nu_k$  & --- &  $G^{\pm}$   \\
				\hline
				\rule[-1ex]{0pt}{2.5ex} 11& VF(\ref{fig:Scoto-Wn}) & $\nu_k$  & --- &   $W^{\pm}$  \\
				\hline \hline
			\end{tabular}
			\caption{Particles involved in each one-loop diagram that contribute to  $h \to l_a^+ l_b^-$ in the Scotogenic model. Second column shows the structure (figure) of each diagram associated to the Figures~\ref{fig:Scoto-Ndiagrams} and~\ref{fig:Scoto-ndiagrams}. The remaining columns identify the particles $P_k$ inside the loop.}
			\label{tab:scoto_contributions}
		\end{table}
	}
\end{center}
As a consequence, the dependence of PV functions is given by 
\begin{subequations}\label{PaVeNu-scoto}
	\begin{align}\nonumber
	\operatorname{C_{0,1,2}} & = \operatorname{C_{0,1,2}}{\left(m_{{N_k}},m_{{\eta}},m_{{\eta}} \right)},\\\nonumber
	\operatorname{B_{1}^{(s)}} & =\operatorname{B_{1}^{(s)}} {\left(m_{{N_k}},m_{{\eta}} \right)},\\
	\nonumber
	\operatorname{B_{0}^{(12)}} & =\operatorname{B_{0}^{(12)}} {\left(m_{{\eta}},m_{{\eta}} \right)},
	\end{align}
\end{subequations}
with $s = 1,2$. Then, the form factors to heavy neutrinos contributions can be calculated with our results as follows.

{\it Diagram~\ref{fig:Scoto-N_etau_etad}:}  FSS structure

This diagram has an FSS structure and two types of couplings are present, which can be decomposed in terms of the coupling constants. Following the structure of our generic vertexes, as given in  equations~\eqref{ec:VerticesGenericos1},~\eqref{ec:VerticesGenericos2}, Figure~\ref{fig:convencionesdiagramas} and the Table~\ref{tab:Vertices_Escoto}, as follows:
\begin{equation}
    \begin{aligned}
    c_h^{S^\pm(1)}(h\eta^- \eta^+) &= -\frac{2i\left(m_{\eta}^{2}-\mu_{2}^{2}\right)}{v}; 
    & &\\
    c_R^{S^+(2)}(\eta^+ l_b^- N_k) &= 0; 
    &c_L^{S^+(2)}(\eta^+ l_b^- N_k) &= i Y_{bk};\\
    c_R^{S^-(3)}(\eta^- l_a^+ N_k) &= i Y_{ak}^{*}; 
    &c_L^{S^-(3)}(\eta^- l_a^+ N_k) &= 0.
\end{aligned}
\end{equation}
Then, by means of equation~\eqref{ec:ALAR-OneFermion} the form factors are given by:
\begin{subequations}\label{ec:ALR_Netaeta}
	\begin{align} \notag
	A_R^h(N_k \eta^{+}\eta^{-})_{ab} &=  c^{S^\pm(1)}_h(h\eta^+ \eta^-) c_L^{S^+(2)}(\eta^+ l_b^- N_k)c_R^{S^-(3)}(\eta^- l_a^+ N_k)\mathcal{H}_{RL}(N_k \eta^{+}\eta^{-})\\\notag
	& = -\frac{2i\left(m_{\eta}^{2}-\mu_{2}^{2}\right)}{v}i Y_{kb}\, i Y_{ka}^{*} \left(\frac{i}{16 \pi^2}(-m_b \operatorname{C}_1) \right)\\
	& = \frac{m_b}{8 \pi^2 v}\left(m_{\eta}^{2}-\mu_{2}^{2}\right) \operatorname{C}_1 Y_{bk}\,  Y_{ak}^{*}, \\\notag
	A_L^h(N_k \eta^{+}\eta^{-})_{ab} 
	&=  c^{S^\pm(1)}_h(h\eta^+ \eta^-) c_L^{S^+(2)}(\eta^+ l_b^- N_k)c_R^{S^-(3)}(\eta^- l_a^+ N_k)\mathcal{H}_{LR}(N_k \eta^{+}\eta^{-}) \\ \notag
	& = -\frac{2i\left(m_{\eta}^{2}-\mu_{2}^{2}\right)}{v}i Y_{bk}\, i Y_{ak}^{*}
	\left(\frac{i}{16 \pi^2}(m_a \operatorname{C}_1) \right) \\
	& = -\frac{m_a}{8 \pi^2 v}\left(m_{\eta}^{2}-\mu_{2}^{2}\right) \operatorname{C}_1  Y_{bk}\,  Y_{ak}^{*}.
	\end{align}
\end{subequations}
We have included the factor $i/16 \pi ^{2}$ which comes from dimensional regularization, for more details see Appendix \ref{appen:loop-identities}. 

\bigskip%

{\it Diagrams~\ref{fig:Scoto-Neta} and~\ref{fig:Scoto-etaN}:} FS and SF structures

For these two diagrams, we only replace $c_h^{S^\pm(1)}(h\eta^+ \eta^-)$ by 
$$c_h^{F^\pm(1)}(h l_a^- l_a^+) = \frac{i g m_a}{2 m_W}$$,
where $m_a$ is the mass of the charged lepton $l_a$. Therefore, by equation~\eqref{ec:ALAR-OneFermion} the form factors for diagrams~\ref{fig:Scoto-Neta} and ~\ref{fig:Scoto-etaN} are
\begin{subequations}\label{ec:ALR_Neta}
	\begin{align}\notag
	A_R^h(N_l \eta)_{ab} &=  m_{ab}^{-2} c^{F^\pm(1)}_h(h l_b^- l_b^+) c_L^{S^+(2)}(\eta^+ l_b^- N_k)c_R^{S^-(3)}(\eta^- l_a^+ N_k)\mathcal{H}_{LR}(N_k \eta) \\ \notag
	& = \frac{i g m_b}{2 m_W m_{ab}^2} i Y_{kb}\,i Y_{ka}^{*} \left(\frac{i}{16 \pi^2}(m_a^2 \operatorname{B}_1^{(1)}) \right)\\ 
	& =  \frac{ g m_a^2 m_b}{32 \pi^2 m_W m_{ab}^2}\operatorname{B}_1^{(1)} Y_{bk}\, Y_{ak}^{*},\\\notag
	A_L^h(N_k \eta)_{ab} &=  m_{ab}^{-2} c^{F^\pm(1)}_h(h l_b^- l_b^+) c_L^{S^+(2)}(\eta^+ l_b^- N_k)c_R^{S^-(3)}(\eta^- l_a^+ N_k)\mathcal{H}_{RL}(N_k \eta)\\\notag
	& =  \frac{i g m_b}{2 m_W m_{ab}^2} i Y_{kb}\,i Y_{ka}^{*} \left(\frac{i}{16 \pi^2}(m_a m_b \operatorname{B}_1^{(1)}) \right)\\ 
	& = \frac{ g m_a m_b^2}{32 \pi^2 m_W m_{ab}^2}\operatorname{B}_1^{(1)} Y_{bk}\, Y_{ak}^{*}, \\ \notag
	A_R^h(\eta N_k)_{ab} &=  m_{ab}^{-2} c^{F^\pm(1)}_h(h l_a^- l_a^+) c_L^{S^+(2)}(\eta^+ l_b^- N_k)c_R^{S^-(3)}(\eta^- l_a^+ N_k)\mathcal{H}_{LR}(\eta N_k) \\ \notag
	& = \frac{i g m_a}{2 m_W m_{ab}^2} i Y_{kb} \, i Y_{ka}^{*} \left(\frac{i}{16 \pi^2}(m_a m_b \operatorname{B}_1^{(2)}) \right)\\ 
	& = \frac{ g m_a^2 m_b}{32 \pi^2 m_W m_{ab}^2}\operatorname{B}_1^{(2)} Y_{bk}\,  Y_{ak}^{*},\\\notag
	A_L^h(\eta N_k)_{ab} &=  m_{ab}^{-2} c^{F^\pm(1)}_h(h l_a^- l_a^+) c_L^{S^+(2)}(\eta^+ l_b^- N_k)c_R^{S^-(3)}(\eta^- l_a^+ N_k)\mathcal{H}_{RL}(\eta N_k)\\\notag
	& =  \frac{i g m_a}{2 m_W m_{ab}^2} i Y_{kb} \, i Y_{ka}^{*}  \left(\frac{i}{16 \pi^2}(m_b^2 \operatorname{B}_1^{(2)}) \right)\\ 
	& = \frac{ g m_a m_b^2}{32 \pi^2 m_W m_{ab}^2}\operatorname{B}_1^{(2)} Y_{bk}\,  Y_{ak}^{*}.
	\end{align}
\end{subequations}
Divergences are canceled after sum the form factors of diagrams~\ref{fig:Scoto-Neta} and~\ref{fig:Scoto-etaN}.
If we denote the loop particles by $\Xi_{N_k} = \{N_k \eta \eta, N_k \eta,\eta N_k\}$, in all the heavy neutrino contributions, the total form factors for heavy neutrinos contribution are as follows:
\begin{align}\notag
A_{L,R}^h(N)_{ab}%
&=\sum_{k=1}^{3}\left( A_{L,R}^h(N_k \eta^{+}\eta^{-})_{ab} + A_{L,R}^h(N_k \eta^{+})_{ab} +A_{L,R}^h(\eta^{+} N_k)_{ab}\right)\\ \label{ec:AnuLR}
&=\sum_{k=1}^{3}\sum_{\xi_k \in \Xi_{N_k}}A_{L,R}^h(\xi_k)_{ab} .
\end{align}
%

\subsubsection{Light neutrino contributions}
\begin{figure}
  \centering
  \begin{subfigure}[t]{.24\linewidth}
    \centering\includegraphics[width=.8\linewidth]{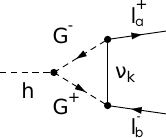}
    \caption{FSS-like}
    \label{fig:Scoto-nGuGd}
  \end{subfigure}
  \begin{subfigure}[t]{.24\linewidth}
    \centering\includegraphics[width=.8\linewidth]{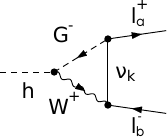}
    \caption{FSV-like}
    \label{fig:Scoto-nGuWd}
  \end{subfigure}
  \begin{subfigure}[t]{.24\linewidth}
    \centering\includegraphics[width=.8\linewidth]{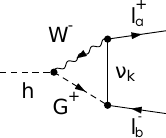}
    \caption{FVS-like}
    \label{fig:Scoto-nWuGd}
  \end{subfigure}
  \begin{subfigure}[t]{.24\linewidth}
    \centering\includegraphics[width=.8\linewidth]{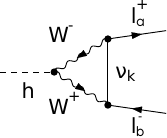}
    \caption{FVV-like}
    \label{fig:Scoto-nWuWd}
  \end{subfigure}
  \medskip
  \begin{subfigure}[t]{.24\linewidth}
    \centering\includegraphics[width=.8\linewidth]{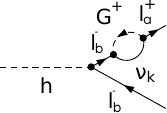}
    \caption{FS-like}
    \label{fig:Scoto-nG}
  \end{subfigure}
  \begin{subfigure}[t]{.24\linewidth}
    \centering\includegraphics[width=.8\linewidth]{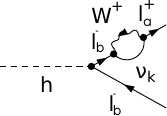}
    \caption{FV-like}
    \label{fig:Scoto-nW}
  \end{subfigure}
  \begin{subfigure}[t]{.24\linewidth}
    \centering\includegraphics[width=.8\linewidth]{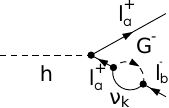}
    \caption{SF-like}
    \label{fig:Scoto-Gn}
  \end{subfigure}
  \begin{subfigure}[t]{.24\linewidth}
    \centering\includegraphics[width=.8\linewidth]{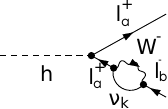}
    \caption{VF-like}
    \label{fig:Scoto-Wn}
  \end{subfigure}
\caption{Light neutrino diagrams contribution to LFV Higgs decays in the Scotogenic model. The sub-captions denote the respective generic structure of each diagram.}
\label{fig:Scoto-ndiagrams}
\end{figure}

Next, we consider the diagrams 4-11 shown in Table~\ref{tab:scoto_contributions} and Figure~\ref{fig:Scoto-ndiagrams}, in this case the particles that contribute to the  loop are Goldstone bosons $G^{\pm}$, vector bosons $W^{\pm}$ and light neutrinos. In Feynman t'Hoft gauge, the Goldstone boson propagator contains $m_{W^{\pm}}$, then Passarino-Veltman functions will have the following mass dependence:
\begin{subequations}\label{PaVenu-scoto}
	\begin{align}\nonumber
	\operatorname{C_{0,1,2}} & = \operatorname{C_{0,1,2}}{\left(m_{{\nu_k}},m_{{W}},m_{{W}} \right)},\\\nonumber
	\operatorname{B_{1}^{(s)}} & =\operatorname{B_{1}^{(s)}} {\left(m_{{\nu_k}},m_{{W}} \right)},\\
	\nonumber
	\operatorname{B_{0}^{(12)}} & =\operatorname{B_{0}^{(12)}} {\left(m_{{W}},m_{{W}}\right)}.
	\end{align}
\end{subequations}

Similarly, as we did for heavy neutrino contributions and following the result of equation~\eqref{ec:ALAR-OneFermion} and Table~\ref{tab:OneFermionContributions}, we obtain the form factors for light neutrino contribution, as given in Table~\ref{tab:AnuLR}.
\begin{table}[h]
	\begin{center}
		\scalebox{0.95}{
			\begin{tabular}{||c|c|c||}
        	\hline \hline
        	\rule[-1ex]{0pt}{4.0ex} {\bf $\xi_k$(Figure)} & $A^h_R(\xi_k)_{ab}$ & $A^h_L(\xi_k)_{ab}$ \\
        	\hline\hline 
        	\rule[-1ex]{0pt}{4.5ex} $\nu_k G^{-} G^{+}$(\ref{fig:Scoto-nGuGd})  & $  -\frac{m_{h}^{2} m_{a}^{2} m_{b} }{8 \pi^{2} v^{3}} \operatorname{C_{1}}
        	 {U}_{bk} {U}_{ak}^*$ 
        	& $\frac{m_{h}^{2} m_{a} m_{b}^{2} }{8 \pi^{2} v^{3}}\operatorname{C_{2}} {U}_{bk} {U}_{ak}^*$ \\
        	\hline
        	\rule[-1ex]{0pt}{4.5ex}  $\nu_k G^{-} W^{+}$(\ref{fig:Scoto-nGuWd}) & $ - \frac{g^{2} m_{a}^2 m_{b} }{32 \pi^{2} v} \left(\operatorname{C_{1}} - 2 \operatorname{C_{2}}\right) {U}_{bk} {U}_{ak}^*$
        	& $\frac{g^{2} m_{a} }{32 \pi^{2} v} \left(m_a^2 \operatorname{C}_1 + 2(m_h^2 - m_a^2)\operatorname{C}_2 -X \right) {U}_{bk} {U}_{ak}^*$ \\
        	\hline
        	\rule[-1ex]{0pt}{4.5ex} $\nu_k W^{-} G^{+}$(\ref{fig:Scoto-nWuGd}) & $-\frac{g^{2} m_{b}}{32 \pi^{2} v} \left(X + m_b^2 \operatorname{C}_2 + 2(m_h^2 - m_b^2)\operatorname{C}_1 \right) {U}_{bk} {U}_{ak}^*$
        	& $ \frac{g^{2} m_{a} m_{b}^2 }{32 \pi^{2} v} \left(\operatorname{C}_2 -2 \operatorname{C}_1\right) {U}_{bk} {U}_{ak}^*$  \\
        	\hline
        	\rule[-1ex]{0pt}{4.5ex} $\nu_k W^{-} W^{+}$(\ref{fig:Scoto-nWuWd})
        	& $ \frac{g^{4} m_{b} v }{32 \pi^{2}} \operatorname{C_{2}} {U}_{bk} {U}_{ak}^*$
        	& $- \frac{g^{4} m_{a} v}{32 \pi^{2}} \operatorname{C_{1}} {U}_{bk} {U}_{ak}^*$  \\
        	\hline
        	\rule[-1ex]{0pt}{4.5ex} $ \nu_k G^{\pm}$(\ref{fig:Scoto-nG})  
        	& $ \frac{\sqrt{2} g m_{a}^{2} m_{b}^{3} }{32 \pi^{2} m_{W} v^{2} m_{ab}^2} \operatorname{{{B^{(1)}_{1}}}} {U}_{bk} {U}_{ak}^*$  
        	& $ \frac{\sqrt{2} g m_{a}^{3} m_{b}^{2} }{32 \pi^{2} m_{W} v^{2} m_{ab}^2} \operatorname{{{B^{(1)}_{1}}}} {U}_{bk} {U}_{ak}^*$ \\
        	\hline
        	\rule[-1ex]{0pt}{4.5ex} $\nu_k W^{\pm}$(\ref{fig:Scoto-nW})  
        	&$ \frac{\sqrt{2} g^{3} m_{a}^{2} m_{b} }{64 \pi^{2} m_{W} m_{ab}^2} \operatorname{{{B^{(1)}_{1}}}} {U}_{bk} {U}_{ak}^*$  
        	& $\frac{\sqrt{2} g^{3} m_{a} m_{b}^{2} }{64 \pi^{2} m_{W} m_{ab}^2} \operatorname{{{B^{(1)}_{1}}}} {U}_{bk} {U}_{ak}^*$  \\
        	\hline
        	\rule[-1ex]{0pt}{4.5ex} $G^{\pm} \nu_k$(\ref{fig:Scoto-Gn})
        	& $  \frac{\sqrt{2} g m_{a}^{2} m_{b}^{3} }{32 \pi^{2} m_{W} v^{2} m_{ab}^2} \operatorname{{{B^{(2)}_{1}}}} {U}_{bk} {U}_{ak}^*$  
        	& $ \frac{\sqrt{2} g m_{a}^{3} m_{b}^{2} }{32 \pi^{2} m_{W} v^{2} m_{ab}^2} \operatorname{{{B^{(2)}_{1}}}} {U}_{bk} {U}_{ak}^*$ \\
        	\hline
        	\rule[-1ex]{0pt}{4.5ex} $W^{\pm} \nu_k$ (\ref{fig:Scoto-Wn})
        	& $  \frac{\sqrt{2} g^{3} m_{a}^{2} m_{b} }{64 \pi^{2} m_{W} m_{ab}^2} \operatorname{{{B^{(2)}_{1}}}} {U}_{bk} {U}_{ak}^*$  
        	& $ \frac{\sqrt{2} g^{3} m_{a} m_{b}^{2} }{64 \pi^{2} m_{W} m_{ab}^2}\operatorname{{{B^{(2)}_{1}}}} {U}_{bk} {U}_{ak}^*$  \\
        	\hline
        	\hline
        \end{tabular}}
	\end{center}
	\caption{Form factors for light neutrino contributions in the Scotogenic model.}
	\label{tab:AnuLR}
\end{table}
Divergences for these form factors are canceled, to see it, we sum over the form factors of  diagrams~\ref{fig:Scoto-nG} with ~\ref{fig:Scoto-Gn}, and~\ref{fig:Scoto-nW} with~\ref{fig:Scoto-Wn}.

Finally, if we denote $\Xi_{\nu_k} = \{\nu_k G^+ G^-, \nu_k G^+ W^-, \nu_k W^+ G^-, \nu_k W^+ W^-, \nu_k G+,G^+ \nu_k, \nu_k W^+, W^+ \nu_k\}$ as the set of all light neutrino contributions, the total form factors for light neutrinos contribution are given by:
\begin{align}\notag
A_{L,R}^h(\nu)_{ab}%
&=\sum_{k=1}^{3}\left( A_{L,R}^h(\nu_k G^+ G^-)_{ab} + A_{L,R}^h(\nu_k G^+
W^-)_{ab} + A_{L,R}^h(\nu_k W^+ G^-)_{ab} + A_{L,R}^h(\nu_k W^+ W^-)_{ab}\right.\\ \notag
& \left.   + A_{L,R}^h(\nu_k G^+)_{ab} + A_{L,R}^h( G^+\nu_k)_{ab} + A_{L,R}^h(\nu_k W^+)_{ab} + A_{L,R}^h(W^+\nu_k)_{ab}\right)\\ 
&=\sum_{k=1}^{3}\sum_{\xi_k \in \Xi_{\nu_k}}A_{L,R}^h(\xi_k)_{ab} .
\label{ec:ANuLR}
\end{align}

\subsection{Numerical analysis of $\mathcal{BR}(h \to l_a l_b)$}
From the results of previous section, we have that light neutrino contribution depends only on the neutrino masses and the mixing matrix $\mathbf{U}^{\nu}$, which is given in~\eqref{ec:Upmns}. Then, considering $m_1= 3.01 \times 10^{-11}$ GeV, the numerical values of the form factors~\eqref{ec:ANuLR} are given in Table~\ref{tab:nu_form_factors}. 
\begin{table}[h]
	\begin{center}
		\begin{tabular}{||c|c|c||}
			\hline \hline
			Decay & $|A_R^{h}(\nu)_{ab}|$ & $|A_L^{h}(\nu)_{ab}|$ \\
			\hline \hline
			$h \to \mu \tau$ & $3.64 \times 10^{-21}$ & $2.17 \times 10^{-22}$ \\
			\hline
			$h \to e \tau$ & $2.32 \times 10^{-20}$ & $1.55 \times 10^{-23}$ \\
			\hline
			$h \to e \mu$  & $6.37 \times 10^{-22}$ & $6.8 \times 10^{-24}$ \\
			\hline \hline
		\end{tabular}
	\end{center}
	\caption{Numerical values of light neutrino contributions to form factors associated to $h \to l_a l_b$ in Scotogenic model.}
	\label{tab:nu_form_factors}
\end{table}

Moreover, heavy neutrino contribution depends on  the masses of $N_k$ and $\eta^{\pm}$, the parameter $\mu_2^2$ and the matrix $\mathbf{Y}$. However, the entries of $\mathbf{Y}$ depend on the parameters $m_1$, $M_k$, $m_R$ and $\lambda_5$, as it is shown in equations~\eqref{ec:Lambda_k} and~\eqref{ec:Y'}\footnote{Formally, $\Lambda_k$ depends on $M_k$, $m_R$ and $m_I$, but we use that $m_I^2 = m_R^2 - 2 \lambda_5 v^2$.}. 
Thus, the allowed parameter, {\it i.e.}, the masses of  $N_1$, $\eta_R$, $\eta^{\pm}$, must be of $O({10^2}) \, \textrm{GeV}$, which is in agreement with previous constraints. We also fix $\mu_2 = 1$ GeV without loss of generality, and $\lambda_5 \approx 10^{-10}$. For heavy neutrinos $N_{1,2,3}$ we consider $M_2 = 10^4\text{ GeV}$, $M_3 = 10^5\text{ GeV}$ and $M_1 \in (10^{2}, 10^{3})\text{ GeV}$.

\begin{figure}
	\centering
	\includegraphics[width=0.8\linewidth]{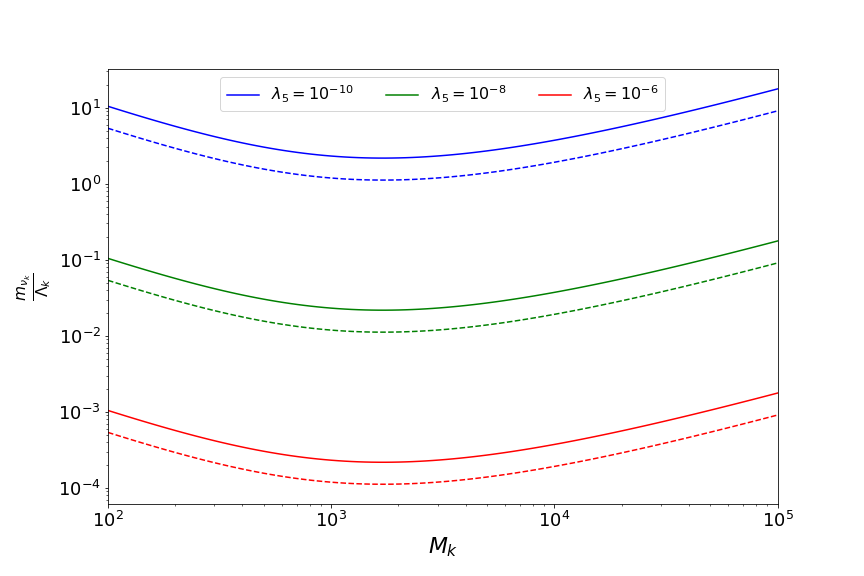}
	\caption{Plot of $m_{\nu_k}/\Lambda_k$ for $k=1$ (dashed lines), $k=3$ (solid lines) and $m_R = 805$ GeV.}
	\label{fig:scotom1l1}
\end{figure}

The form factors of the first three diagrams in Table \ref{tab:scoto_contributions} with heavy neutrinos are proportional to $Y_{kb}Y_{ka}^{*}$ from equations \eqref{ec:ALR_Netaeta} and \eqref{ec:ALR_Neta}. Also, from~\eqref{ec:YdagaY} we know that $Y_{kb}Y_{ka}^{*} \propto m_{\nu_k}/\Lambda_k$~\footnote{The case for $k=2$ is similar to case $k=1$.}. We derive that $ m_{\nu_k}/\Lambda_k$ increases like $\lambda_{5}$ decreases as illustrated in Figure~\ref{fig:scotom1l1}. Then, $\lambda_5 \approx 10^{-10}$ implies large values of $m_{\nu_1}/\Lambda_1$, and as a consequence, large heavy neutrino contributions to $\mathcal{BR}(h \to l_a l_b)$.

\begin{figure}
	\centering
    \centering\includegraphics[width=0.8\linewidth]{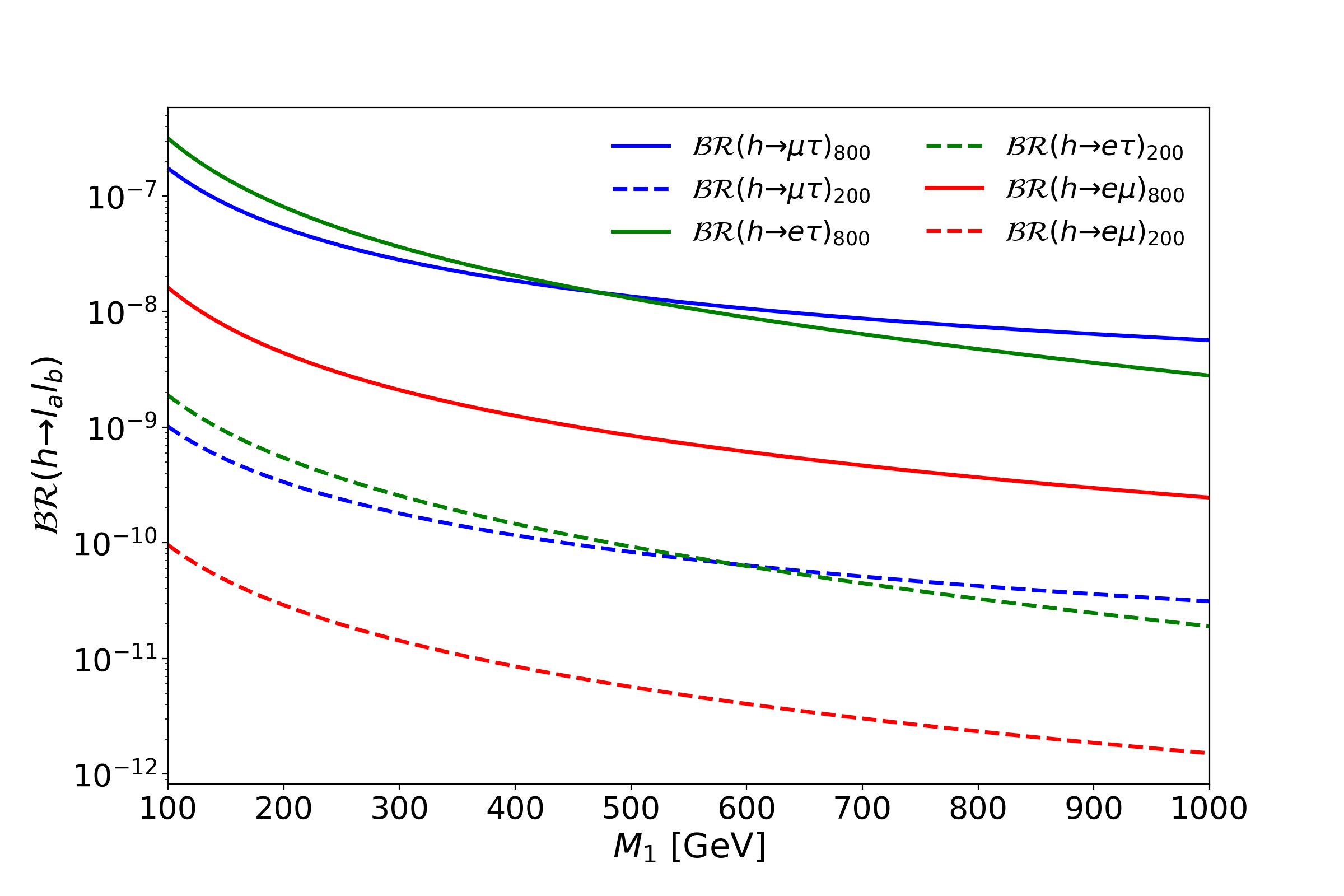}
	\caption{Numerical results for $\br(h \rightarrow l_a l_b)$ vs $M_1$ in the Scotogenic model for $M_{2} = 10^{4}$ GeV and $M_3 = 10^5$ GeV, $\lambda_5 = 1 \times 10^{-10}$, $\mu_2 = 1$ GeV. The plot shows two cases: $(i)$ $\br(h \to l_a l_b)_{800}$ with $m_\eta = 800$ GeV and $m_R = 805$ GeV (solid lines); $(ii)$ $\br(h \to l_a l_b)_{200}$ with $m_\eta = 200$ GeV and $m_R = 205$ GeV (dashed lines).}
	\label{fig:BR-scoto-no-nu}
\end{figure}

Figure~\ref{fig:BR-scoto-no-nu} shows the $\mathcal{BR}(h \to l_a l_b)$ for the Scotogenic model for two scenarios: $(i)$ for $m_\eta = 800$ GeV, $m_R = 805$ GeV (solid lines); and $(ii)$ for $m_\eta = 200$ GeV, $m_R = 205$ GeV (dashed lines). The largest values of $\mathcal{BR}(h \to l_a l_b)$ are reached for $M_1 = 100$ GeV, which are: $\mathcal{BR}(h \to \mu \tau)_{800} \approx \mathcal{BR}(h \to e \tau)_{800} \approx 10^{-7}$ and $\mathcal{BR}(h \to e \mu)_{800} \approx 10^{-8}$, also $\mathcal{BR}(h \to \mu \tau)_{200} \approx \mathcal{BR}(h \to e \tau)_{200} \approx 10^{-9}$ and $\mathcal{BR}(h \to e \mu)_{200} \approx 10^{-10}$.   

\begin{figure}
	\centering
    \includegraphics[width=0.9\linewidth]{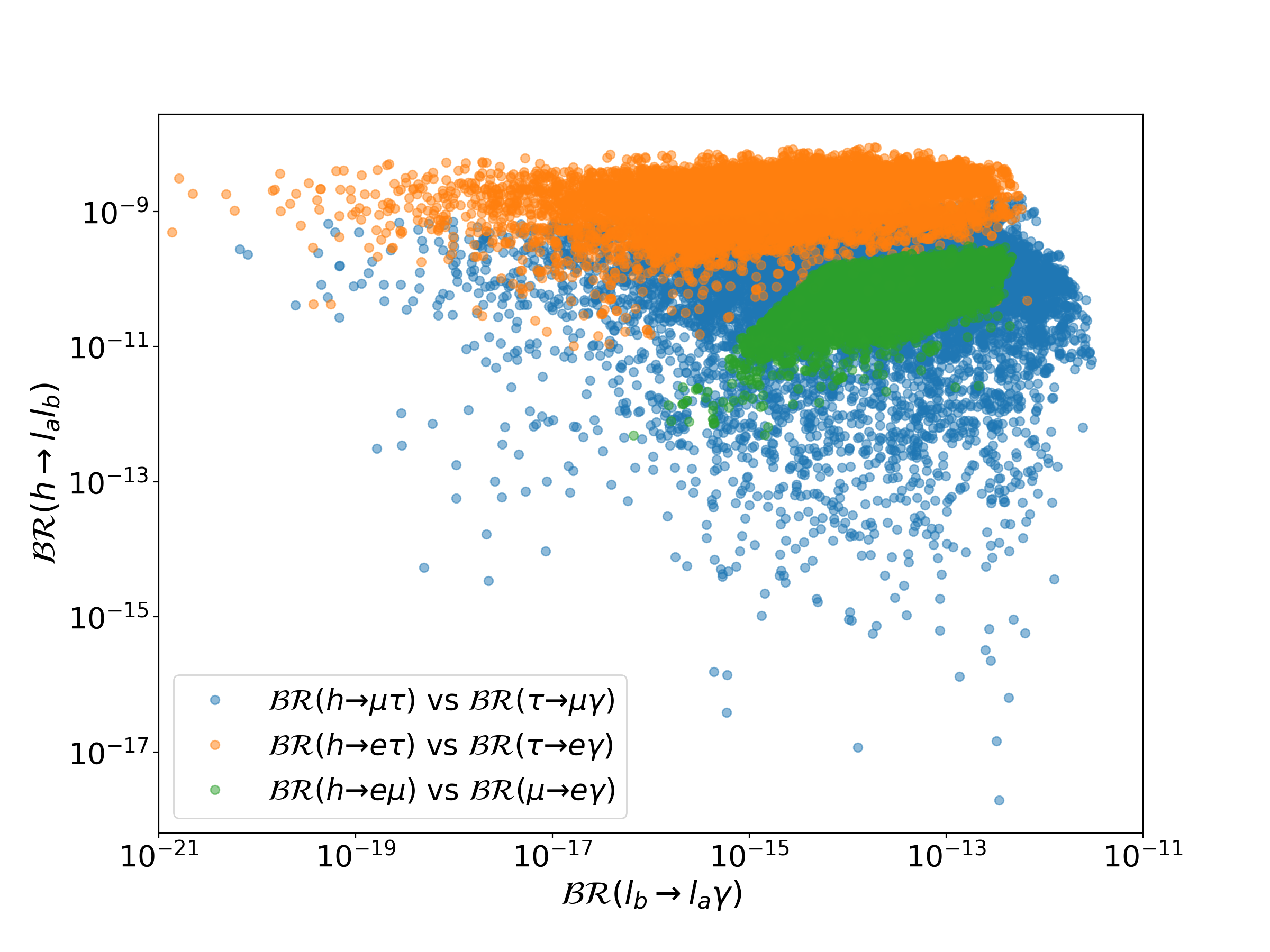}
	\caption{Comparison of $\mathcal{BR}(h \rightarrow l_a l_b)$ and $\mathcal{BR}(l_b \rightarrow l_a \gamma)$  in the allowed space in the Scotogenic model.}
	\label{fig:BRhliljvsBljligamma}
\end{figure}

In Ref.~\cite{Herrero-Garcia2016} they found that $\br(h \to \mu \tau) \lessapprox 10^{-7} \lambda_3^2$. In this work we write $\lambda_3 = \frac{2}{v^2}(m_\eta^2 - \mu_2^2)$ in terms of $m_\eta \approx 10^{2}$ GeV and $\mu_2=1$ GeV. Then by means of equation~\eqref{ec:masas-escalares}, the corresponding bound of~\cite{Herrero-Garcia2016} will be given by $\br(h \to \mu \tau) \lessapprox 10^{-7}$, which is in agreement with our results. Recently, the work~\cite{Hundi:2022iva} attempted to find a enhance in LFV Higgs decays signals in the Scotogenic model, they assume $\lambda_3 \approx 1$, and $\lambda_5 <<1$ which implies $Y_{{ab}}\approx 1$. In our scan, we obtain $m_{\eta} \approx 10^{2}$ and $\lambda_5 \approx 10^{-10}$ which implies $\lambda_3 \approx 1$ and $Y_{{ab}}\approx 1$, naturally, we also agree with these results.

Finally, we have previously analyzed the consequences of electroweak parameters, CLFV process and relic density of dark matter bounds over the parameter space. However, we found that $\mathcal{BR}(h \rightarrow l_a l_b )\lesssim 10^{-9}$, as it is shown in Figure~\ref{fig:BRhliljvsBljligamma}. These results satisfy the limits reported by ATLAS and BABAR given in Table~\ref{tb:bounds-lfv}.

\section{Conclusions\label{sec:concl}}

In this paper we have studied the  general formulae for the calculation of one-loop LFV Higgs decays $H_r \to l_a l_b$, with $H_r$ being part of the Higgs spectrum of a multi-scalar extension of the  SM that includes neutrino masses. In particular, we calculated the form factors for all diagrams that contribute to the LFV Higgs decays, where FCNC are forbidden at tree level; the structure of these diagrams depend on the number of fermions inside the loop. The results calculated in Section~\ref{sec:higgsint} have been implemented in the code named \verb|OneLoopLFVHD| through the \verb|Python| library for symbolic calculations \href{https://www.sympy.org}{Sympy}, which enhances its usefulness in the \href{https://jupyter.org/}{Jupyter notebook} graphical interface. With the help of \href{https://mpmath.org/}{mpmath} the symbolic expressions can be implemented numerically with arbitrary precision. The code and examples can be downloaded in the GitHub repository \href{https://github.com/moiseszeleny/OneLoopLFVHD}{OneLoopLFVHD}. 

Then, we applied our general formulae for two particular models, and we found agreement  for See-Saw Type I-$\nu SM$  with the results given in~\cite{PILAFTSIS199268,PhysRevD.47.1080,PhysRevD.71.035011,THAO2017159}. Finally, for this model, we found that the largest values are $\mathcal{BR}(h\to \mu \tau)\simeq 10^{-12}$ for $m_{n_6} \approx 10^{15}$ GeV. 

In the case of the Scotogenic model, we found a region of parameter space  that satisfies all the constraints from LFV process, dark matter relic density, electroweak precision tests and perturvativity. For LFV Higgs decays, we separate the contributions of heavy and light neutrinos, and we found that the largest contributions come from the heavy neutrinos and light neutrino contribution is negligible. We identify an upper bound on the branching ratios of the order $\mathcal{BR}(h\to l_a l_b) \lesssim 10^{-9}$ for the Scotogenic model. 
On the other hand, for the heavier Higgs bosons, the corresponding branching ratio could reach much larger values; but we shall present these results elsewhere.

{\centering {\bf Acknowledgments}}

M. Z. M. thanks to Conacyt (Mexico) for the PhD fellowship.  J. L. D. C. and O. F. B.  thank the support of SNI (Mexico).

\appendix
\section{Generic Feynman Rules\label{appen:feynmanrules}}
\begin{figure}[h]
	\begin{center} 
	\begin{subfigure}[t]{.3\linewidth}
    \includegraphics[width=\linewidth]{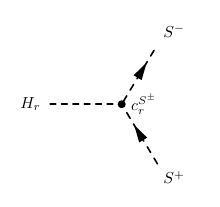}
    \caption{Interaction $H_r S^{-} S^{+}$}
    \end{subfigure}
    \begin{subfigure}[t]{.3\linewidth}
	\includegraphics[width=\linewidth]{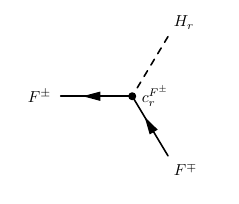}
    \caption{Interaction $H_r F^{\pm} F^{\mp}$}
    \end{subfigure}
    \begin{subfigure}[t]{.3\linewidth}
		\includegraphics[width=\linewidth]{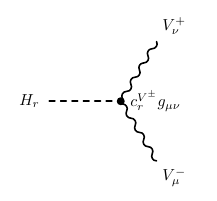}
    \caption{Interaction $H_r V^{+} V^{-}$}
    \end{subfigure}
    \begin{subfigure}[t]{.3\linewidth}
		\includegraphics[width=\linewidth]{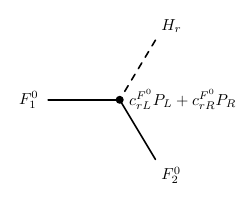}
    \caption{Interaction $H_r F_1^0 F_2^0$}
    \end{subfigure}
    \begin{subfigure}[t]{.3\linewidth}
		\includegraphics[width=\linewidth]{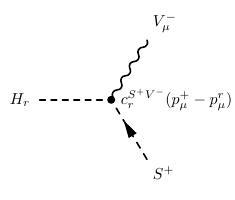}
    	\caption{Interaction $H_r V^- S^+$}
    \end{subfigure}
    \begin{subfigure}[t]{.3\linewidth}
		\includegraphics[width=\linewidth]{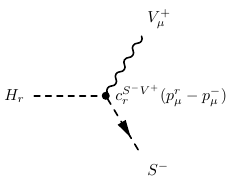}
        \caption{Interaction $H_r V^+ S^-$}
    \end{subfigure}
\end{center}
\caption{Feynman rules for scalar, fermionic and vector interactions for $H_r$.}
\label{Feynrules:Ha}
\end{figure}

\begin{figure}[h]
	\begin{center}
		\begin{subfigure}[t]{.4\linewidth}
			\includegraphics[width=\linewidth]{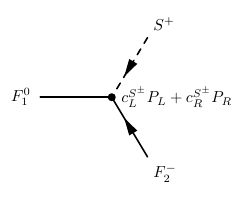}
			\caption{Interaction $S^+ F_1^0 F_2^-$}
		\end{subfigure}
		\begin{subfigure}[t]{.44\linewidth}
			\includegraphics[width=\linewidth]{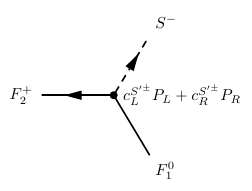}
			\caption{Interaction $S^- F_1^0 F_2^+$}
		\end{subfigure}
		\begin{subfigure}[t]{.43\linewidth}
			\includegraphics[width=\linewidth]{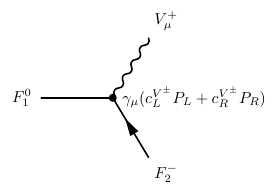}
			\caption{Interaction $V^+ F_1^0 F_2^-$}
		\end{subfigure}
		\begin{subfigure}[t]{.44\linewidth}
			\includegraphics[width=\linewidth]{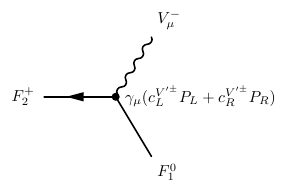}
			\caption{Interaction $V^- F_1^0 F_2^+$}
		\end{subfigure}
	\end{center}
	\caption{Feynman rules for  charged current interactions mediated by charged scalars and vectors.}
	\label{Feynrules:Charged}
\end{figure}

The Feynman rules for the models under consideration, have a common structure, as given in Figures~\ref{Feynrules:Ha} and~\ref{Feynrules:Charged}. First, we have the scalar, fermionic and vector interactions with $H_r$ as it is shown in Figure~\ref{Feynrules:Ha}.
For models where FCNC are forbidden at tree level, LFV process are at loop-level mediated only by charged currents either fermionic or bosonic. The generic Feynman rules for these interactions are given in Figure~\ref{Feynrules:Charged}.
%

\section{Loop formulae and identities\label{sec:aploopfor}\label{appen:loop-identities}}
We use the conventions given in~\cite{Hue:2015fbb,THAO2017159} for the Passarino-Veltman functions, and using those mass assignments shown in Figure~\ref{fig:convencionesdiagramas}, denominators of propagators are given by
\begin{equation}\label{ec:PropD012}
\operatorname{D}_0 = k^2 - M_0^2 + i \delta, \quad \operatorname{D}_1 = (k-p_1)^2 - M_1^2 + i \delta, \quad  \operatorname{D}_2 = (k + p_2)^2 - M_2^2 + i \delta,
\end{equation}
where $\delta$ is an infinitesimal parameter that allows to perform the loop integrals correctly; $M_i$ denotes the mass of particle $P_i$ inside the loop, with $p_1$, $p_2$ denoting the lepton momentum  $l_a$, $l_b$, respectively.

We use dimensional regularization, where the four dimensional momentum integral can be rewritten as
\begin{equation}\label{ec:intetracionD}
\int \frac{d^4 k}{(2 \pi)^4} \rightarrow \frac{i}{16 \pi^2}\times \frac{(2\pi \mu)^{4-D}}{i \pi^2}\int d^4k,
\end{equation}
where $\mu$ is a parameter with mass dimensions. Following the conventions of~\cite{Hue:2015fbb}, this step will be omitted, the final results will be obtained by including the  factor $i/16\pi^2$. 

In the Feynman gauge, only the following scalar integrals are present:
\begin{subequations}
	\begin{eqnarray}
	\operatorname{B}_0^{(i)}(M_0, M_i) = N_D \int\frac{d^{D}k}{\operatorname{D}_0 \operatorname{D}_{i}},  &\operatorname{B}_0^{(12)}(M_1, M_2) = N_D \int\frac{d^{D}k}{\operatorname{D}_1 \operatorname{D}_2},\\ 
	  \operatorname{C}_0(M_0,M_1,M_2) = \frac{1}{i \pi^2}
	 \int\frac{d^{4}k}{\operatorname{D}_0 \operatorname{D}_1 \operatorname{D}_2},&
	\end{eqnarray}
\end{subequations}
where $N_D = (2\pi \mu)^{4-D}/i\pi^2$ and $D = 4-2\epsilon$. The tensor integrals are the following
\begin{subequations}
	\begin{align} 
	\operatorname{B}^{\mu}(p_i, M_0, M_i) &= N_D \int\frac{d^{D}k\times k^\mu}{D_0 D_i} = \operatorname{B}_1^{(i)}p_i^{\mu},\\ 
	\operatorname{C}^\mu = \operatorname{C}^{\mu}(M_0,M_1,M_2) &= \frac{1}{i \pi^2} \int\frac{d^{4}k\times k^\mu}{D_0 D_1 D_2} = \operatorname{C}_1 p_1^{\mu} + \operatorname{C}_2 p_2^{\mu}.
	\end{align}
\end{subequations}
On the other hand, we know that $\operatorname{C}_i$ ($i = 0,1,2$) integrals are finite, but the $\operatorname{B}$ functions contain divergences. The divergent integrals are written in terms of its divergent and finite terms (denoted by the lowercase):
\begin{subequations}\label{ec:PaVetoDivFin}
	\begin{align}
	\operatorname{B}_{0,1}^{(i)} &= Div[\operatorname{B}_{0,1}^{(i)}] + b_{0,1}^{(i)},\\
	\operatorname{B}_{0}^{(12)} &= Div[\operatorname{B}_{0}^{(12)}] + b_{0}^{(12)}.
	\end{align}
\end{subequations}
Also,
\begin{subequations}\label{ec:PaVeDiv}
	\begin{align}
	Div[\operatorname{B}_0^{(i)}] &= Div[\operatorname{B}_0^{(12)}] = \Delta_\epsilon,\\
	Div[\operatorname{B}_1^{(1)}] &= - Div[\operatorname{B}_1^{(2)}] = \frac{1}{2}\Delta_\epsilon,
	\end{align}
\end{subequations}
where $\Delta_\epsilon = 1/\epsilon + \ln{4\pi} - \gamma_E + \ln{\frac{\mu^2}{m_r^2}}$ with $\gamma_E$ the Euler constant, and $m_r$ is the mass $H_r$. For simplicity in calculation, we use approximate expressions for PV functions, where $p_1^2, \, p_2^2 \to 0$. The $\operatorname{C}_0$ function is given by~\cite{Hue:2015fbb,Bardin:1999ak}:
\begin{equation}\label{ec:C0}
\operatorname{C}_{0}=\frac{1}{m_{r}^{2}}\left[R_{0}\left(x_{0}, x_{1}\right)+R_{0}\left(x_{0}, x_{2}\right)-R_{0}\left(x_{0}, x_{3}\right)\right],
\end{equation}
where 
\begin{equation}\label{ec:R0}
R_{0}\left(x_{0}, x_{i}\right) \equiv L i_{2}\left(\frac{x_{0}}{x_{0}-x_{i}}\right)-L i_{2}\left(\frac{x_{0}-1}{x_{0}-x_{i}}\right),
\end{equation}
with $Li_2$ the dilogarithm function. Also,
\begin{equation}\label{ec:x0x3}
x_{0}=\frac{M_{2}^{2}-M_{0}^{2}}{m_{r}^{2}}, \quad x_{3}=\frac{-M_{0}^{2}+i \delta}{M_{1}^{2}-M_{0}^{2}},
\end{equation}
and the $x_{1,2}$ are solutions of 
\begin{equation}\label{ec:solx12}
x^{2}-\left(\frac{m_{r}^{2}-M_{1}^{2}+M_{2}^{2}}{m_{r}^{2}}\right) x+\frac{M_{2}^{2}-i \delta}{m_{r}^{2}}=0.
\end{equation}
The finite part of $\operatorname{B}^{(i)}_{0,1}$ functions can be evaluated in a numerically stable way using~\cite{DENNER200662} 
\begin{subequations}\label{ec:B_Denner}
	\begin{align}
	b_{0}^{(i)}&=-\ln{ M_{0}^{2}}-\sum_{l=1}^{2} f_{0}\left(y_{kl}\right), \,  \\
	b_{1}^{(i)}&=(-1)^{i-1} \frac{1}{2}\left(-\ln{M_{0}^{2}} - \sum_{l=1}^{2} f_{1}\left(y_{kl}\right)\right),
	\end{align}
\end{subequations}
where $y_{kl}$ with $k = i,j$, $l=1,2$ are solutions of 
\begin{subequations}
\begin{align}
y^{2} p_i^2 - y(p_i^2 + M_0^2 - M_1^2) + M_0^2 -i \delta &= 0,\\
y^{2} p_j^2 - y(p_j^2 + M_0^2 - M_2^2) + M_0^2 -i \delta &= 0,
\end{align}
\end{subequations} 
and are given by 

\begin{subequations}
	\begin{align}
	y_{i} &= \frac{1}{2 p_i^2}\left[p_i^2 + M_0^2 - M_1^2 \pm \sqrt{(p_i^2 + M_0^2 - M_1^2)^2 - 4 p_i^2 M_0^2} \right], \\
	y_{j} &= \frac{1}{2 p_j^2}\left[p_j^2 + M_0^2 - M_2^2 \pm \sqrt{(p_j^2 + M_0^2 - M_2^2)^2 - 4 p_j^2 M_0^2} \right].
	\end{align}
\end{subequations}
Also,
\begin{subequations}\label{ec:f01}
	\begin{align}
	f_0(y) &= y \ln(-y) - y \ln(1-y) + \ln(1-\frac{1}{y}) - 1\\
	f_1(y) &= y^2\ln(-y) -y^2 \ln(1-y) -y + \ln(1-\frac{1}{y}) - \frac{1}{2},
	\end{align}
\end{subequations}
derived from
\begin{equation}\label{ec:fn}
f_{n}(x) \equiv(n+1) \int_{0}^{1} \mathrm{~d} t \, t^{n} \ln \left(1-\frac{t}{x}\right).
\end{equation}
These functions have poles in $y =0,1$.

The expressions for $b_0^{(12)}$ and $\operatorname{C}_i$ are as follows~\cite{THAO2017159}
\begin{subequations}\label{ec:PaVeFin}
	\begin{align}
	b_0^{(12)}&= -\ln{M_1^2} + 2 + \sum_{k=1}^{2}x_k \ln{\left(1 -\frac{1}{x_k} \right) },\\
	\operatorname{C}_1 &= \frac{1}{m_r^2} \left[ b_0^{(1)} - b_0^{(12)} + (M_2^2 - M_0^2) \operatorname{C}_0\right],\\
	\operatorname{C}_2 &= -\frac{1}{m_r^2} \left[ b_0^{(2)} - b_0^{(12)} + (M_1^2 - M_0^2) \operatorname{C}_0\right],
	\end{align}
\end{subequations}
where $b_0^{(12)}$ depends on $x_k$, which are solutions of equation~\eqref{ec:solx12}. 
If we focus on the definition of $b_0^{(12)}$, we observe that this definition has poles when $x_k = 0,1$. From equation~\eqref{ec:solx12}, $x_k = x_k(M_1,M_2,m_r)$ and we can deduce the following limit cases:
	\begin{subequations}\label{ec:x12_anomalies}
		\begin{eqnarray}
		x_{1,2} &= 1 \text{ for } M_1 &= 0 \text{ and } M_2 = m_r, \\
		x_{1,2} &= 0 \text{ for } M_1 &= m_r \text{ and } M_2 = 0.
		\end{eqnarray}
	\end{subequations}
The  expressions~\eqref{ec:C0} and~\eqref{ec:PaVeFin} have been taken from~\cite{Hue:2015fbb} and numerically evaluated, the results are compared with LoopTools~\cite{10.1093/ptep/ptw158} finding a good agreement.

\section{Cancellation of divergences in the $\nu$SM \label{Appendix:Cancellation_divergences}}
It follows from equations~\eqref{ec:seesaw-mass} and ~\eqref{ec:diagonalization} that
\begin{equation}\label{ec:seesaw-indentities}
\begin{aligned}
M_{a b}^{\nu}=0, &\quad M_{(I+3)(J+3)}^{\nu}=\left(M_{N}\right)_{I J}, \quad M_{a(I+3)}^{\nu}=\left(M_{D}\right)_{a I}, \quad M_{(I+3) a}^{\nu}=\left(M_{D}^{T}\right)_{I a}, \\
\mathbf{U}^{\nu \dagger} \mathbf{U}^{\nu}=\mathbf{I}, & \quad \mathbf{M}^{\nu}=\mathbf{U}^{\nu *} \hat{\mathbf{M}}^{\nu} \mathbf{U}^{\nu \dagger}, \quad \text{and} \quad \mathbf{M}^{\nu *}=\mathbf{U}^{\nu} \hat{\mathbf{M}}^{\nu} \mathbf{U}^{\nu T}.
\end{aligned}
\end{equation}
These relations will be useful in the cancellation of divergences form the form factors.
In particular, to analyze the divergent terms we focus forts for the $R$-form factors $A_R$, while
those of the $L$-factors can be done along the same lines. All diagrams in the $\nu$SM are given in Table~\ref{tab:seesaw_contributions}, we focus only on diagrams 1, 7, 8, 9 and 10, which contain divergent terms.

Let us start with the divergences for diagram 1 given by
\begin{align}\notag
Div\left[A_R^{(1)}\right]& = {m}_{b}\sum_{i,j = 1}^{6}   \left[\Delta_\epsilon m_{n_i}^{2} {{{C}}}_{ij} + \Delta_\epsilon  m_{n_i} m_{n_j}C^*_{ij} 
\right] \Delta_{ij}^{ab}\\ \notag
& = {m}_{b}\sum_{i,j = 1}^{6}   \left[ \Delta_\epsilon m_{n_i}^{2} {{{C}}}_{ij}\right] \Delta_{ij}^{ab}.
\end{align}
Note that in second line, we have considered
\begin{align}\notag
\sum_{i,j = 1}^{6} m_{n_i} m_{n_j}C^*_{ij} \Delta_{ij}^{ab} &= \frac{g^{3}}{64 \pi^{2} m_{W}^{3}}\sum_{i,j = 1}^{6}\sum_{c=1}^{3} m_{n_i} m_{n_j}U_{ci}^{\nu *} U_{cj}^{\nu} U_{bj}^{\nu} U_{ai}^{\nu *}\\ \notag
& = \frac{g^{3}}{64 \pi^{2} m_{W}^{3}}\sum_{j = 1}^{6}\sum_{c=1}^{3} m_{n_j}U_{bj}^{\nu}U_{cj}^{\nu}    \sum_{i = 1}^{6} m_{n_i} U_{ci}^{\nu *}  U_{ai}^{\nu *}\\ \label{ec:id-mni-mnj}
& = \frac{g^{3}}{64 \pi^{2} m_{W}^{3}}\sum_{j = 1}^{6}\sum_{c=1}^{3}m_{n_j}U_{bj}^{\nu}U_{cj}^{\nu} (U^{\nu * } \hat{M}_\nu, U^{\nu \dagger})_{ac} = 0,
\end{align}
and definitions of $C_{ij}$ and $\Delta_{ij}^{ab}$~\eqref{ec:Dijab}. On the other hand,
\begin{align}\notag
\sum_{i,j = 1}^{6}  m_{n_{i}}^2{C}_{ij} \Delta_{ij}^{ab} &= \frac{g^{3}}{64 \pi^{2} m_{W}^{3}}\sum_{i,j = 1}^{6}\sum_{c=1}^{3} m_{n_{i}}^2U_{ci}^{\nu} U_{cj}^{\nu *} U_{bj}^{\nu} U_{ai}^{\nu *}\\ \notag
& = \frac{g^{3}}{64 \pi^{2} m_{W}^{3}}\sum_{i = 1}^{6}\sum_{c=1}^{3} m_{n_{i}}^2 U_{ci}^{\nu} U_{ai}^{\nu *}\sum_{j = 1}^{6}U_{bj}^{\nu}U_{cj}^{\nu *}\\ \notag
& = \frac{g^{3}}{64 \pi^{2} m_{W}^{3}}\sum_{i = 1}^{6}\sum_{c=1}^{3} m_{n_{i}}^2 U_{ci}^{\nu} U_{ai}^{\nu *}(U^{\nu}U^{\nu \dagger})_{bc}\\\label{ec:id-mni2}
& = \sum_{i = 1}^{6}m_{n_{i}}^2 \Delta_{ii}^{ab},
\end{align}
where unitarity of $U^{\nu}$ has been used. Considering equations~\eqref{ec:id-mni-mnj} and~\eqref{ec:id-mni2}, the divergent term of $A_R^{(1)}$ is 
\begin{align}\notag
Div\left[A_R^{(1)}\right] &= {m}_{b}\Delta_\epsilon\sum_{i = 1}^{6}  m_{n_i}^{2} \Delta_{ii}^{ab}.
\end{align}

Now, consider the diagrams 8 and 10, its right divergent terms are given by 
\begin{align}\notag
Div\left[A_R^{(8)}\right] & =  \frac{{m}_{b} }{{m}_{a}^{2} - {m}_{b}^{2}}\sum_{i=1}^{6}\left( m_{n_i}^2 \left({m}_{a}^{2} + {m}_{b}^{2}\right) \Delta_\epsilon - m_a^2 \left({m}_{b}^{2} + m_{n_i}^{2}\right)\frac{\Delta_\epsilon}{2} \right)\Delta_{ii}^{ab}\\
& = \frac{ {m}_{b} }{{m}_{a}^{2} - {m}_{b}^{2}}\sum_{i=1}^{6}\left( \frac{ \Delta_\epsilon}{2} m_{n_i}^2 m_a^2 +m_b^2 \Delta_\epsilon \left(m_{n_i}^2 - \frac{m_a^2}{2} \right) \right)\Delta_{ii}^{ab},
\end{align}
\begin{align}\notag
Div\left[A_R^{(10)}\right] & = - \frac{ {m}_{a}^2 m_b}{{m}_{a}^{2} - {m}_{b}^{2}}\sum_{i=1}^{6}\left(\left({m}_{b}^{2} + m_{n_i}^{2}\right) \left(- \frac{\Delta_\epsilon}{2}\right ) + 2 \Delta_\epsilon m_{n_i}^{2} \right)\Delta_{ii}^{ab} \\
& = -\frac{ {m}_{a}^2 m_b}{{m}_{a}^{2} - {m}_{b}^{2}}\sum_{i=1}^{6} \left( \frac{3 \Delta_\epsilon}{2} m_{n_i}^2 - m_b^2 \frac{ \Delta_\epsilon}{2}  \right)\Delta_{ii}^{ab}.
\end{align}
After some simplifications we obtain $$Div\left[A_R^{(8)}\right] + Div\left[A_R^{(10)}\right]  = - {m}_{b}\Delta_\epsilon\sum_{i = 1}^{6}  m_{n_i}^{2} \Delta_{ii}^{ab},$$ then divergent terms of diagrams 1, 8 and 10 disappear when we sum each other.

In the case of diagrams 7 and 9 we have the next divergent terms 
\begin{align}
Div\left[A_R^{(7)}\right] & = -  \frac{2 m_{W}^{2}  {m}_{a}^2 {m}_{b} }{{m}_{a}^{2} - {m}_{b}^{2}}\sum_{i=1}^{6}\frac{\Delta_\epsilon}{2}\Delta_{ii}^{ab},
\end{align}
\begin{align}
Div\left[A_R^{(9)}\right] & =  - \frac{2 m_{W}^{2}  {m}_{a}^{2} {m}_{b} }{ {m}_{a}^{2} - {m}_{b}^{2}}\sum_{i=1}^{6}\left(-\frac{\Delta_\epsilon}{2}\right)\Delta_{ii}^{ab},
\end{align}
but each of them is null by the GIM mechanism.
\bibliographystyle{plain}

\end{document}